\documentclass[aps,prb,superscriptaddress,showpacs,twocolumn,bibnotes]{revtex4}
\usepackage{graphicx}
\usepackage{amssymb}
\usepackage{amsmath}

\newcommand{\bk}{{\bf k}}
\newcommand{\bq}{{\bf q}}
\newcommand{\bqD}{{\bf q}_{\mathrm D}}

\newcommand{\bK}{{\bf K}}

\newcommand{\bp}{{\bf p}}

\newcommand{\br}{{\bf r}}

\newcommand{\bJ}{{\bf J}}

\newcommand{\bdelta}{{\boldsymbol\delta}}
\newcommand{\bsigma}{{\boldsymbol\sigma}}

\newcommand{\sgn}{{\mathop{\rm{sgn}}\nolimits\,}}

\newcommand{\Tr}{{\mathop{\rm{Tr}}\nolimits\,}}

\newcommand{\vF}{v_{\mathrm{F}}}

\begin{document}

\title{Transport properties of graphene across strain-induced nonuniform
velocity profiles}

\author{F. M. D. Pellegrino}
\affiliation{Dipartimento di Fisica e Astronomia, Universit\`a di Catania,\\
Via S. Sofia, 64, I-95123 Catania, Italy}
\affiliation{CNISM, UdR Catania, I-95123 Catania, Italy}
\author{G. G. N. Angilella}
\affiliation{Dipartimento di Fisica e Astronomia, Universit\`a di Catania,\\
Via S. Sofia, 64, I-95123 Catania, Italy}
\affiliation{CNISM, UdR Catania, I-95123 Catania, Italy}
\affiliation{Scuola Superiore di Catania, Universit\`a di Catania,\\
Via Valdisavoia, 9, I-95123 Catania, Italy}
\affiliation{INFN, Sez. Catania, I-95123 Catania, Italy}
\author{R. Pucci}
\affiliation{Dipartimento di Fisica e Astronomia, Universit\`a di Catania,\\
Via S. Sofia, 64, I-95123 Catania, Italy}
\affiliation{CNISM, UdR Catania, I-95123 Catania, Italy}

\date{\today}

\begin{abstract}
We consider the effect of uniaxial strain on ballistic transport in graphene,
across single and multiple tunneling barriers. Specifically, we show that
applied strain not only shifts the position of the Dirac points in reciprocal
space, but also induces a deformation of the Dirac cones, and that both effects
are of the same order on the applied strain intensity. We therefore study the
deviations thereby induced on the angular dependence of the tunneling
transmission across a single barrier, as well as on the conductivity and Fano
factor across a single barrier and a superstructure of several, periodically
repeated, such sharp barriers. Our model is generalized to the case of
nonuniform barriers, where either the strain or the gate potential profiles may
depend continuously on position. This should afford a more accurate description
of realistic `origami' nanodevices based on graphene, where `foldings' are
expected to involve several lattice spacings.
\medskip
\end{abstract} 

\maketitle

\section{Introduction}

Graphene is an atomically thin, two-dimensional layer of carbon atoms arranged
according to a honeycomb lattice. After having being speculated since long as
the ideal building block of graphite and other $sp^2$ carbon compounds, it has
been recently obtained in the laboratory \cite{Novoselov:05a}, thereby kindling
an extraordinary outburst of experimental as well as theoretical research
activity \cite{CastroNeto:08,Abergel:10}. Reduced dimensionality and its
peculiar structure conspire towards the formation of low-energy quasiparticles,
which can be described as massless Dirac fermions with a cone dispersion
relation in reciprocal space around the so-called Dirac points $\bK$,
$\bK^\prime$, and a linearly vanishing density of states (DOS) at the Fermi
level. This is reflected in several electronic properties already in the
non-interacting limit, \emph{e.g.} Klein tunneling
\cite{MiltonPereira:06,Barbier:09,Barbier:10,Barbier:10a,Peres:09}, the
reflectivity \cite{Nair:08}, the optical conductivity
\cite{Kuzmenko:08,Wang:08,Mak:08,Stauber:08a,Pellegrino:09b}, and the plasmon
dispersion relation \cite{Hwang:07a,Polini:09,Pellegrino:10a,Pellegrino:10c}.

Graphene is also remarkable for its exceptional mechanical properties, as is
generic for most carbon compounds. For instance, notwithstanding its reduced
dimensionality, graphene is characterized by a relatively large tensile strength
and stiffness \cite{Booth:08}, with graphene sheets being capable to sustain
elastic deformations as large as $\approx 20$\%
\cite{Kim:09,Liu:07,Cadelano:09,Choi:10,Jiang:10}. Larger strains would then
induce a semimetal-to-semiconductor transition, with the opening of an energy
gap \cite{Gui:08,Pereira:08a,Ribeiro:09,Cocco:10}, and it has been demonstrated
that such an effect critically depends on the direction of applied strain
\cite{Pellegrino:09b,Pellegrino:09c}. The effect of uniaxial strain on the
linear response electronic properties of graphene has been studied on quite
general grounds \cite{Pellegrino:11}.

Recently, it has been suggested that graphene-based electronic devices might be
designed by suitably tailoring the electronic structure of a graphene sheet
under applied strain \cite{Pereira:09}. Indeed, a considerable amount of work
has been devoted to the study of the transport properties in graphene across
strain-induced single and multiple barriers \cite{Cayssol:09,Gattenloehner:10}.
There, the main effect of strain has usually been considered to be that of
shifting the position of the Dirac points in reciprocal space. However, it has
been demonstrated that a nonuniform space variation of the underlying gate
potential would result in a modulation of the Fermi velocity
\cite{Cayssol:09,Concha:10,Raoux:10}.

Here, we show that both effects are of the same order on the applied strain
intensity, and should therefore be considered on the same ground, when studying
the transport properties of strained graphene. We shall therefore explicitly
consider not only the strain-induced displacement of the Dirac points in
reciprocal space, but also a strain-induced deformation of the Dirac cones,
resulting in a strain-dependent anisotropic Fermi velocity. Specifically, we
will consider tunneling through a single strain-induced sharp barrier, possibly
subjected to a gate potential, and through a superstructure made of several such
barriers, periodically repeated. More interestingly, we will generalize our
results to the problem of transport through a tunneling structure, characterized
by a \emph{nonuniform} variation of both the Fermi velocity and of the gate
potential, as can \emph{e.g.} be brought about by a continuous deformation or
applied uniaxial strain.

The paper is organized as follows. After introducing our model in
Sec.~\ref{sec:model}, we discuss the effect of a strain-induced modulation of
the Fermi velocity on the angular dependence of the transmission across a single
sharp barrier, as well as on the conductivity and Fano factor for ballistic
transport (Sec.~\ref{sec:single}). We then consider the case of several such
barriers, arranged in a periodic fashion (Sec.~\ref{sec:multiple}). In
Sec.~\ref{sec:smooth}, we generalize our results to the case of
\emph{nonuniform} strain across a smooth barrier. Finally, in
Sec.~\ref{sec:conclusions} we summarize and give directions for future
investigation.

\section{Model}
\label{sec:model}

In unstrained graphene, low-energy quasiparticles can be described by the linear
Hamiltonian in momentum space
\begin{equation}
H^{(0)} = \hbar \vF {\mathbb{I}} \bsigma \cdot \bp,
\label{eq:Hunstrained}
\end{equation}
where $\vF$ is the Fermi velocity, $\bsigma = (\sigma_1 , \sigma_2 )$, with
$\sigma_i$ and $\tau_i$ ($i=1,2,3$) Pauli matrices and ${\mathbb{I}}$ the
identity matrix associated with the two-dimensional spaces of the sublattices
($A$ and $B$, say), and of the two valleys around the Dirac points ($\bK$ and
$\bK^\prime$), respectively. Eq.~(\ref{eq:Hunstrained}) acts on the
four-component spinors \cite{Aleiner:06,Basko:08a}
\begin{equation}
\Psi_\bp = (\Psi_{A,\bK} (\bp) , \Psi_{B,\bK} (\bp) , \Psi_{B,\bK^\prime} (\bp)
, -\Psi_{A,\bK^\prime} (\bp))^\top,
\label{eq:4spinor}
\end{equation}
where $\bp$ is measured from the Dirac point one is referring to. Here and
below, a superscript zero denotes absence of strain. The effect of uniaxial
strain in real space is that of modifying the lattice vectors as $\bdelta_\ell =
(\mathbb{I} +  {\boldsymbol\varepsilon}) \cdot \bdelta^{(0)}_\ell$
($\ell=1,2,3$), where $\bdelta_1^{(0)} = {\mathfrak{a}}(\sqrt{3},1)/2$, 
$\bdelta_2^{(0)} = {\mathfrak{a}}(-\sqrt{3},1)/2$,  $\bdelta_3^{(0)} =
{\mathfrak{a}}(0,-1)$ are the relaxed (unstrained) vectors connecting two
nearest-neighbor (NN) carbon sites, with ${\mathfrak{a}}=1.42$~\AA, the
equilibrium C--C distance in a graphene sheet \cite{CastroNeto:08}, and
${\boldsymbol\varepsilon}$ is the strain tensor \cite{Pereira:08a}
\begin{equation}
{\boldsymbol\varepsilon} = 
\frac{1}{2} \varepsilon [(1-\nu){\mathbb I} + (1+\nu) A(\theta)],
\label{eq:strainmat}
\end{equation}
where
\begin{equation}
A (\theta) 
= \cos(2\theta) \sigma_z + \sin(2\theta) \sigma_x ,
\end{equation}
where the Pauli matrices now are understood to act on vectors of the
two-dimensional direct or reciprocal lattice. In Eq.~(\ref{eq:strainmat}),
$\theta$ denotes the angle along which the strain is applied, with respect to
the $x$ axis in the lattice coordinate system, $\varepsilon$ is the strain
modulus, and $\nu$ is Poisson's ratio. While in the hydrostatic limit $\nu=-1$
and ${\boldsymbol\varepsilon} = \varepsilon{\mathbb I}$, in the case of graphene
one has $\nu=0.14$, as determined from \emph{ab initio} calculations
\cite{Farjam:09}, to be compared with the known experimental value $\nu=0.165$
for graphite \cite{Blakslee:70}. The special values $\theta=0$ and
$\theta=\pi/2$ refer to strain along the zig~zag and armchair directions,
respectively.

The possibility of describing the effects of strain through
Eq.~(\ref{eq:strainmat}), \emph{i.e.} elastically, implies that applied strain
does not induce any irreversible process or mechanical failure of the graphene
sheet, such as dislocations, grain boundaries, or cracks. In fact, such dramatic
effects are not expected for strain below $\sim 20$~\%, as is predicted by
calculations within density functional theory \cite{Liu:07,Marianetti:10}, and
confirmed experimentally by means of atomic force microscopy (AFM)
\cite{Lee:08}.

In momentum space, the effect of uniaxial strain on the Hamiltonian
Eq.~(\ref{eq:Hunstrained}) is likewise accounted for by the strain tensor,
Eq.~(\ref{eq:strainmat}). This is usually described as a shift in momentum space
of the location of the Dirac points. However, starting from the more general,
tight-binding Hamiltonian \cite{CastroNeto:08}, expanding to first order in the
strain modulus, and to second order in the impulses, one may show that applied
strain also induces a deformation of the Dirac cones, at the same (first) order
in $\varepsilon$. Explicitly, one finds
\begin{widetext}
\begin{eqnarray}
H &=& 
\hbar \vF \sigma_1 \left[
\left( 1 + \left( \frac{1}{2} - \kappa_0 \right) \varepsilon (1-\nu)
+ \left( \frac{1}{2} - \frac{1}{2} \kappa_0 \right) \varepsilon (1+\nu)
\cos(2\theta) \right) p_x + \left( \frac{1}{2} - \frac{1}{2} \kappa_0 \right)
\varepsilon (1+\nu) \sin (2\theta) p_y \right] \nonumber \\
&&+ \hbar \vF \sigma_2 \left[
\left( 1 + \left( \frac{1}{2} - \kappa_0 \right) \varepsilon (1-\nu)
- \left( \frac{1}{2} - \frac{1}{2} \kappa_0 \right) \varepsilon (1+\nu)
\cos(2\theta) \right) p_y + \left( \frac{1}{2} - \frac{1}{2} \kappa_0 \right)
\varepsilon (1+\nu) \sin (2\theta) p_x \right] \nonumber \\
&&-
\frac{1}{4} \hbar \vF \tau_3 {}
\left[ \sigma_1 (p_x^2 - p_y^2 ) -2\sigma_2 p_x p_y \right] 
- \hbar\vF \tau_3 \sigma_1 \varepsilon (1+\nu)\cos(2\theta) 
+ \hbar\vF \tau_3 \sigma_2 \varepsilon (1+\nu)\sin(2\theta),
\label{eq:Hamexpl}
\end{eqnarray}
\end{widetext}
where $\kappa_0 = ({\mathfrak{a}}/2t)|\partial t/\partial {\mathfrak{a}}|
\approx 1.6$ is related to the logarithmic derivative of the nearest-neighbor
hopping $t$ at $\varepsilon=0$. 

Our model is based on the tight-binding approximation for the band structure,
including only nearest-neighbor hopping. To this level of approximation, one
does not observe any strain-induced modification of the work function, $\Phi$.
In order to include also such effects, one needs to consider also next-nearest
neighbor hopping \cite{CastroNeto:08}. Making use of the expression for the
hopping function between two neighboring carbon $p$-orbitals involved in a $\pi$
bond, as a function of the bond length $\ell$, $V_{pp\pi} (\ell) = t_0 e^{-3.37
(\ell/{\frak a} -1)}$, with $t_0 = -2.7$~eV \cite{Pereira:08a}, one finds
\begin{equation} 
\Phi = \frac{3}{2} (1-\nu) \sqrt{3} {\frak a} \left.
\frac{dV_{pp\pi} (\ell)}{d\ell} \right|_{\ell = \sqrt{3} {\frak a}} \varepsilon
\approx 1.7
\mathrm{~eV} \times \varepsilon , 
\label{eq:work} 
\end{equation} 
\emph{viz.} a scalar term, going linear with the strain modulus $\varepsilon$,
whose order of magnitude agrees with the \emph{ab initio} results of
Ref.~\onlinecite{Choi:10}. At any rate, the work function, Eq.~(\ref{eq:work}),
can be absorbed in an effective scalar potential $U$, which we conventionally
refer to as to a gate potential below.

Another effect that is not explicitly considered in our model is the deformation
of the $\pi$ orbitals due to off-plane bending, as would be \emph{e.g.}
generated by an AFM tip. However, a change in the hopping parameters due to the
bending of the graphene sheet can be described as an effective in-plane strain
\cite{Kim:08}. Specifically, one may expect that strain induced by an AFM tip
would be characterized by cilindrical symmetry, which is beyond the scope of the
present work, where only linear barriers are considered. We note in passing that
other efficient ways to realize controllable strain consists in depositing
graphene on top of deformable substrates \cite{Bao:09,Mohiuddin:09}.

The spectrum of the strained Hamiltonian, Eq.~(\ref{eq:Hamexpl}), is still
linear, but now around the shifted Dirac points $\bqD {\mathfrak{a}} = \pm
(\kappa_0 \varepsilon(1+\nu)\cos(2\theta), -\kappa_0
\varepsilon(1+\nu)\sin(2\theta))^\top$. To first order in the wavevector
displacement $\bq = \bp \mp \bqD$ from such shifted Dirac points, one finds
\begin{equation}
H=\hbar\vF \bsigma\cdot\bq^\prime ,
\end{equation}
where
\begin{equation}
\bq^\prime =
[(1-\kappa\varepsilon(1-\nu)) {\mathbb I} -\kappa\varepsilon(1+\nu)A(\theta)]
\bq ,
\end{equation}
and $\kappa=\kappa_0 - \frac{1}{2}$. However, it is convenient to work in the
reference frame with the $x$ axis along the direction of applied strain. This is
accomplished by a rotation in the  sublattice $AB$ space, described by the
unitary matrix
\begin{equation}
U(\theta) = \begin{pmatrix} 1 & 0 \\ 0 & e^{-i\theta} \end{pmatrix} ,
\label{eq:rotation}
\end{equation}
so that
\begin{equation}
H = \hbar\vF U^\dag (\theta) \left[ \sigma_1 (1-\lambda_x \varepsilon ) q_x +
\sigma_2 (1-\lambda_y \varepsilon) q_y \right] U(\theta) ,
\label{eq:Hrot}
\end{equation}
where $\lambda_x = 2\kappa$, $\lambda_y = -2\kappa\nu$. After the rotation,
Eq.~(\ref{eq:rotation}), the location of the Dirac points is given by
\begin{equation}
\bqD {\mathfrak{a}} = \pm (\kappa_0
\varepsilon(1+\nu)\cos(3\theta), -\kappa_0
\varepsilon(1+\nu)\sin(3\theta))^\top .
\label{eq:Diracrot}
\end{equation}
The density operator can be expressed as
\begin{equation}
\rho(\br) = \Psi^\dag (\br) \Psi(\br) ,
\end{equation}
where $\Psi(\br) = (2\pi)^{-2} \int d^2 \bk e^{-i\bk\cdot\br} \Psi (\bk)$.
Correspondingly, the current density operator can be derived as \cite{Paul:03}
$\bJ = -\frac{ie}{\hbar}[H,\br]$, yielding
\begin{equation}
J_i (\br) = -e\vF \Psi^\dag (\br) (1-\lambda_i \varepsilon) U^\dag (\theta)
\sigma_i U(\theta) \Psi(\br).
\label{eq:J}
\end{equation}
In the following, for the sake of definitiveness, we shall restrict to the
valley $\bK$ only, thus having $\bq=\bp-\bqD$.

\section{Tunneling across a single barrier}
\label{sec:single}

\begin{figure}[t]
\centering
\includegraphics[width=\columnwidth]{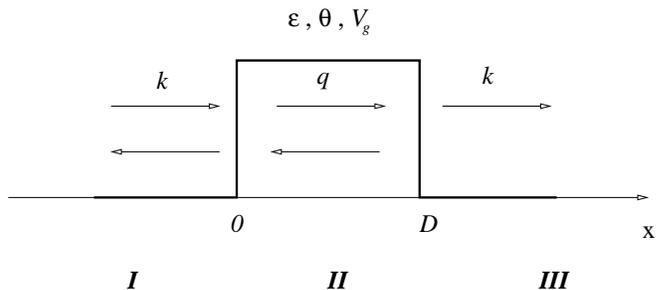}
\caption{One-dimensional single tunneling barrier along the $x$ direction.
Region II ($0\leq x \leq D$) is characterized by applied strain $\varepsilon$
along the $\theta$ direction, as well as by a gate voltage $V_g$.}
\label{fig:barrier}
\end{figure}

Potential barriers for single quasiparticle tunneling in graphene are
conventionally designed by suitably changing the underlying gate voltage.
Recently, it has been suggested that an equivalent effect may be induced by
local uniaxial strain \cite{Kim:08,Pereira:09}. Therefore, we start by
considering a strain-induced one-dimensional step-like barrier, characterized by
uniaxial strain applied along the direction $\theta$, with respect to the $x$
axis, Eq.~(\ref{eq:strainmat}), with strain modulus $\varepsilon$ for $0\leq
x\leq D$, and zero otherwise. Correspondingly, the Hamiltonian and current
density vector are given by Eqs.~(\ref{eq:Hrot}) and (\ref{eq:J}), respectively.
In addition, for the sake of generality, we may also consider a nonzero gate
potential $V_g$ within the barrier (Fig.~\ref{fig:barrier}).

Since we are interested in stationary solutions and the strain-barrier is
uniform along the $y$ direction, the energy $E$ and the component $k_y$ of the
wavevector of an incoming wave are conserved. We look therefore for solutions of
the stationary Dirac equation of the form
\begin{equation}
\psi(x,y) = \left\{
\begin{matrix}
U^\dag (\theta) \psi_{\mathrm{I}} (x) e^{ik_y y}, & \quad x < 0 ,\\
U^\dag (\theta) \psi_{\mathrm{II}} (x) e^{ik_y y}, & \quad 0 \leq x \leq D ,\\
U^\dag (\theta) \psi_{\mathrm{III}} (x) e^{ik_y y}, & \quad x > D ,
\end{matrix}
\right.
\end{equation}
where
\begin{subequations}
\label{eq:singlesols}
\begin{eqnarray}
\label{eq:singlesols:I}
\psi_{\mathrm{I}} (x) &=&
\left[
\frac{1}{\sqrt{2}} 
\begin{pmatrix} 1 \\ s e^{i\varphi} \end{pmatrix}
e^{ik_x x} \right.\nonumber\\
&&+\left.
\frac{r}{\sqrt{2}} 
\begin{pmatrix} 1 \\ - s e^{-i\varphi} \end{pmatrix}
e^{-i k_x x}
\right]  
, \\
\label{eq:singlesols:II}
\psi_{\mathrm{II}} (x) &=&
\left[
\frac{a}{\sqrt{2}} 
\begin{pmatrix} 1 \\ s^\prime e^{i\alpha} \end{pmatrix}
e^{i(q_x +q_{\mathrm{D}}) x} \right.\nonumber\\
&&+ \left.
\frac{b}{\sqrt{2}} 
\begin{pmatrix} 1 \\ - s^\prime e^{-i\alpha} \end{pmatrix}
e^{-i(q_x - q_{\mathrm{D}} )x }
\right] 
, \\
\label{eq:singlesols:III}
\psi_{\mathrm{III}} (x) &=&
t 
\begin{pmatrix} 1 \\ s e^{i\varphi} \end{pmatrix}
e^{ik_x x}
.
\end{eqnarray}
\end{subequations}
In Eqs.~(\ref{eq:singlesols}), $\varphi$ denotes the angle of incidence with
respect to the barrier, $k_x = (|E|/\hbar\vF) \cos\varphi$, $k_y =
(|E|/\hbar\vF) \sin\varphi$, $(E-U_g)^2 = \hbar^2\vF^2 [ (1-\lambda_x
\varepsilon)^2 q_x^2 + (1-\lambda_y \varepsilon)^2 (k_y - q_{{\mathrm{D}}y})^2
]$, $s^\prime = \sgn (E-U_g )$, with $U_g = -eV_g$ Propagating waves correspond
to real values of $q_x$, while evanescent waves correspond to having $q_x$
purely immaginary.

Given the stationary character of the solution, the continuity equation implies
that $\nabla\cdot\bJ = 0$ everywhere. In particular,
$\langle\bJ\rangle\equiv\langle\psi|\bJ|\psi\rangle$ may only depend on $x$,
therefore $\langle J_x \rangle$ is constant. The latter condition implies, at
the barrier boundaries,
\begin{subequations}
\label{eq:continuity}
\begin{eqnarray}
\psi_{\mathrm{I}} (0^-) &=& (1-\lambda_x \varepsilon)^{-1/2} \psi_{\mathrm{II}}
(0^+), \\
(1-\lambda_x \varepsilon)^{-1/2} \psi_{\mathrm{II}} (D^-) &=&
\psi_{\mathrm{III}} (D^+) .
\end{eqnarray}
\end{subequations}
Enforcing the above conditions in Eqs.~(\ref{eq:singlesols}), one eventually
finds for the tunneling transmission, $T=|t|^2$, 
\begin{equation}
T =  \frac{C^2 \cos^2 \varphi}{C^2\cos^2\varphi\cos^2(q_x D)
+(1-ss^\prime S\sin\varphi)^2 \sin^2 (q_x D)},
\label{eq:Tpropagating}
\end{equation}
where $q_y = k_y - q_{\mathrm{D}y}$, $q_x = (1-\lambda_x \varepsilon)^{-1}
|(E-U_g)^2/\hbar^2\vF^2 -  (1-\lambda_y \varepsilon)^2 q_y^2|^{1/2}$, 
$C = (1-\lambda_x \varepsilon) \hbar\vF q_x / |E-U_g |$, 
$S = (1-\lambda_y \varepsilon) \hbar\vF q_y / |E-U_g |$.

\subsection{Angular dependence}

\begin{figure}[t]
\centering
\includegraphics[height=\columnwidth,angle=-90]{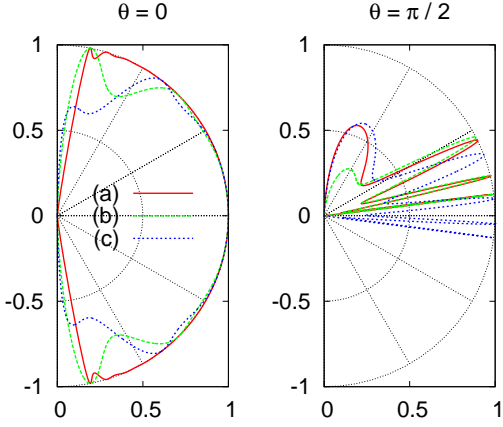}
\caption{(Color online) Dependence on the incidence angle $\varphi$ of the
tunneling transmission $T$, Eq.~(\ref{eq:Tpropagating}). Left panel refers to
strain applied along the zig~zag direction ($\theta=0$), and (a)
$\varepsilon=0.03$, $U_g = 0$~meV; (b) $\varepsilon=0.03$, $U_g = -20$~meV (the
strain-induced deformation of the Dirac cone is neglected); (c)
$\varepsilon=0.03$, $U_g = -20$~meV. Right panel refers to strain applied along
the armchair direction ($\theta=\pi/2$), and (a) $\varepsilon=0.01$, $U_g =
0$~meV; (b) $\varepsilon=0.01$, $U_g = 0$~meV (the strain-induced deformation of
the Dirac cone is neglected); (c) $\varepsilon=0.01$, $U_g = -20$~meV.}
\label{fig:singletrans80}
\end{figure}

\begin{figure}[t]
\centering
\includegraphics[height=\columnwidth,angle=-90]{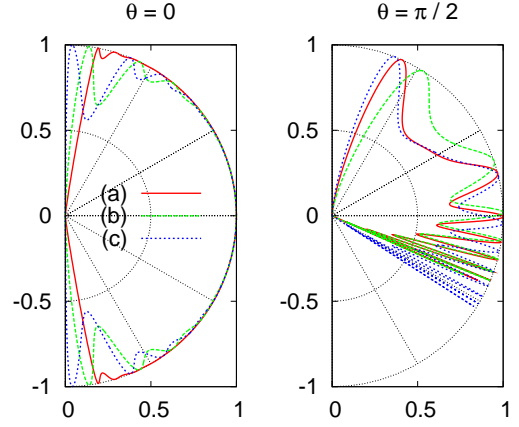}
\caption{(Color online) Same as Fig.~\ref{fig:singletrans80}, but for
$E=150$~meV and $D=100$~nm.}
\label{fig:singletrans150}
\end{figure}

In order to discuss the dependence of the tunneling transmission on the
incidence angle $\varphi$, we preliminarly observe that propagation within the
barrier is allowed whenever
\begin{equation}
\hbar^2 \vF^2 (1-\lambda_y \varepsilon)^2 (k_y - q_{\mathrm{D}y})^2 \leq (E-U_g
)^2 ,
\label{eq:propcond}
\end{equation}
where $k_y = (E/\hbar\vF ) \sin\varphi$. Within such a range, one has moreover
total transmission ($T=1$) whenever
\begin{equation}
q_x D = n\pi ,
\label{eq:peaks}
\end{equation}
$n$ being an integer. Eq.~(\ref{eq:propcond}) differs from the usual condition
for propagation across strain-induced barriers \cite{Pereira:09} in that we are
not only considering a shift of the Dirac point $\bqD$, but also a
strain-induced deformation of the Dirac cone, here exemplified by the
substitution $\vF \mapsto \vF (1-\lambda_y \varepsilon)$.

Figs.~\ref{fig:singletrans80} and \ref{fig:singletrans150} show our results for
the tunneling transmission $T=T(\varphi)$, Eq.~(\ref{eq:Tpropagating}), as a
function of the incidence angle $\varphi$, for $E=80$~meV, $D=100$~nm
(Fig.~\ref{fig:singletrans80}) and $E=150$~meV, $D=100$~nm
(Fig.~\ref{fig:singletrans150}). In both figures, left (\emph{resp.,} right)
panel refers to uniaxial strain applied along the zig~zag ($\theta=0$;
\emph{resp.,} armchair, $\theta=\pi/2$) direction. 

In the case of strain applied along the zig~zag direction ($\theta=0$,
Figs.~\ref{fig:singletrans80} and \ref{fig:singletrans150}, left panels), curves
(b) neglect a strain-induced deformation of the Dirac cone. Comparison with
curves (c), where such a deformation is fully included, shows that the effect of
a strain-induced anisotropy of the Fermi velocity is that of shifting the
angular location of the maxima ($T=1$, Eq.~(\ref{eq:peaks})) of the tunneling
transmission. Such an effect becomes more important with increasing energy (from
Fig.~\ref{fig:singletrans80} to Fig.~\ref{fig:singletrans150}), while the number
of peaks increases, Eq.~(\ref{eq:peaks}), and the angular range in which the
propagating regime is allowed widens. The effect of a strain-induced deformation
of the Dirac cone is even more dramatic in the absence of a gate potential [$U_g
= 0$~meV, curve (a)]. Indeed, in such a case, neglecting the Fermi velocity
anisotropy for strain applied along the zig~zag direction would yield a uniform
tunneling transmission $T=1$, for all incidence angles $\varphi$, whereas we
find that transmission via propagating waves is allowed only for $|\varphi| \leq
\arcsin [(1-\lambda_y \varepsilon)^{-1}]$, with small oscillations below $T=1$
within, and evanescent waves beyond that range. A similar analysis applies to
the case of strain applied along the armchair direction ($\theta=\pi/2$,
Figs.~\ref{fig:singletrans80} and \ref{fig:singletrans150}, right panels), which
is characterized by an asymmetric transmission $T=T(\varphi)$, with 
pronounced oscillations for $\varphi>0$ close to the propagating edge.

The origin of such an asymmetry of the $\varphi$-dependence of the transmission
can be traced back to the particular Dirac cone vertex $\bqD$, whose shift is
here considered. Global symmetry would be restored upon inclusion of the other
Dirac cone. In that case, one would obtain the same picture, but with
$\varphi\mapsto - \varphi$. It should be emphasized that the stationariety
condition, Eq.~(\ref{eq:peaks}), characterizes the occurrence of peaks in the
transmission $T(\varphi)$ in any case. In addition, for a potential barrier, in
the absence of strain, one also recovers complete transmission ($T=1$) at
$\varphi=0$ (Klein tunneling).

Summarizing, at variance with previous studies \cite{Pereira:09}, from
Eq.~(\ref{eq:Tpropagating}) one obtains that the overall effect of a
strain-induced deformation of the Dirac cones is that of shifting the
transmission peaks, and of reducing the range in $\varphi$ at which transmission
takes place.

\subsection{Ballistic transport}
\label{sec:ballistic}

\begin{figure}[t]
\centering
\includegraphics[width=\columnwidth]{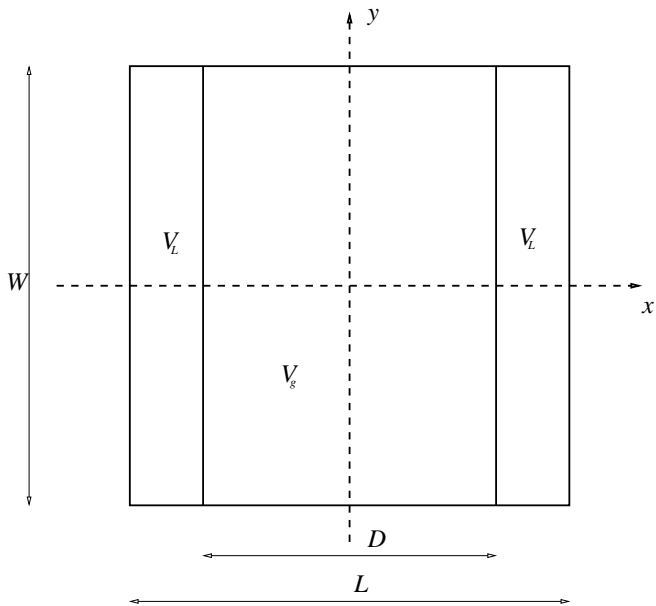}
\caption{Schematic top view of a graphene layer contacted by metallic leads,
as considered in Sec.~\ref{sec:ballistic}.}
\label{fig:strip}
\end{figure}

We now consider a more realistic device, \emph{viz.} a graphene strip of length
$L$ and width $W$, subjected to two leads at a distance $D$
(Fig.~\ref{fig:strip}) \cite{Tworzydlo:06,Gattenloehner:10,Concha:10,Hannes:10}.
Following Ref.~\onlinecite{Tworzydlo:06}, we assume that $W/L\gg 1$, and that
the gate potential within the strip is much less than the potential of the
leads, $|V_g |\ll |V_L |$. Moreover, we assume that the graphene strip be
characterized by uniaxial strain, with modulus $\varepsilon$ and strain
direction $\theta$, and explicitly consider the deformation of the Dirac cones
induced by the applied strain. The energy levels of
Ref.~\onlinecite{Tworzydlo:06} are therefore modified into
\begin{subequations}
\begin{eqnarray}
E &=& U_L + s\hbar\vF \sqrt{k_x^2 + k_y^2} , \quad x<0, ~~ x > D,\\
&=& U_g + s^\prime \hbar\vF \sqrt{(1-\lambda_x \varepsilon)^2 q_x^2 +
(1-\lambda_y \varepsilon)^2 q_y^2} , \nonumber\\
&& \qquad\qquad\qquad 0 < x < D,
\end{eqnarray}
\end{subequations}
where again $q_y = k_y - q_{\mathrm{D}y}$, and $U_L = -eV_L$ and $U_g = -eV_g$.
The limit $|V_L |\to\infty$ is equivalent to the limit $\varphi\to0$, and the
transmission, Eq.~(\ref{eq:Tpropagating}), reduces to
\begin{equation}
T_\alpha^{\mathrm{prop}} (k_y ) = \frac{1}{\cos^2 (q_x D) + \eta(k_y ) \sin^2
(q_x D)},
\end{equation}
for propagating waves in the valley $\alpha=\bK$, where
\begin{equation}
\eta(k_y ) = \frac{(E-U_g )^2}{(E-U_g )^2 - \hbar^2 \vF^2 (1-\lambda_y
\varepsilon)^2 (k_y \pm q_{\mathrm{D}y})^2} ,
\end{equation}
and the minus (\emph{resp.,} plus) sign applies to the valley $\alpha=\bK$
(\emph{resp.,} $\alpha=\bK^\prime$). Analogous expressions hold for the
transmission $T_\alpha^{\mathrm{evan}} (k_y )$ in the evanescent case, with
$\eta(k_y ) \mapsto -\eta(k_y)$, $\cos(q_x D) \mapsto \cosh (q_x D)$, and
$\sin(q_x D) \mapsto \sinh (q_x D)$. The transmission for a general (propagating
or evanescent) wave therefore reads
\begin{equation}
T_\alpha (k_y ) = \Theta[\eta(k_y)] T_\alpha^{\mathrm{prop}} (k_y ) +
(1-\Theta[\eta(k_y)]) T_\alpha^{\mathrm{evan}} (k_y ) ,
\end{equation}
where $\Theta(t)$ is the Heaviside (step) function. Integrating over $k_y$ and
summing over both valleys, one obtains the conductance across the barrier
(Landauer formula) \cite{Landauer:57,Buettiker:86}
\begin{equation}
G = \frac{2e^2}{h} W \sum_\alpha \int_{-\infty}^\infty
\frac{dk_y}{2\pi} T_\alpha (k_y ),
\label{eq:Landauer1}
\end{equation}
where the factor of 2 takes into account for the spin degeneracy, the conductivity
\begin{equation}
\sigma = \frac{D}{W} G,
\label{eq:Landauer}
\end{equation}
and the Fano factor \cite{Fano:57,Blanter:00}
\begin{equation}
F = 1 - \frac{\sum_\alpha \int_{-\infty}^\infty
\frac{dk_y}{2\pi} T_\alpha^2 (k_y )}{\sum_\alpha \int_{-\infty}^\infty
\frac{dk_y}{2\pi} T_\alpha (k_y )} .
\label{eq:Fano}
\end{equation}
In Eq.~(\ref{eq:Landauer}) for the conductivity, the summation over the valleys
contributes with an additional factor of two, whereas this factor cancels in
the definition of the Fano factor, Eq.~(\ref{eq:Fano}).

Before discussing our results, let us observe that the inclusion of a
strain-induced deformation of the Dirac cone in the expressions of the
conductivity, Eq.~(\ref{eq:Landauer}), and of the Fano factor, Eq.~(\ref{eq:Fano}),
amounts to the replacements
\begin{subequations}
\begin{eqnarray}
D &\mapsto& D_{\mathrm{eff}} \equiv \xi D,\\
E &\mapsto& E_{\mathrm{eff}} \equiv \zeta E,
\end{eqnarray}
\end{subequations}
for the strip width and incident energy, respectively, in the
corresponding expressions, $\sigma^{(0)}$ and $F^{(0)}$, say, without cone
deformation, with
\begin{subequations}
\label{eq:eff}
\begin{eqnarray}
\xi &=& \frac{1-\lambda_y \varepsilon}{1-\lambda_x \varepsilon} ,\\
\zeta &=& \frac{1}{1-\lambda_y \varepsilon} .
\end{eqnarray}
\end{subequations}
In particular, one explicitly finds
\begin{equation}
\sigma(D,E) = \xi^{-1} \sigma^{(0)} (D_{\mathrm{eff}},
E_{\mathrm{eff}}) .
\label{eq:sigmaeff}
\end{equation}
As a consequence, while $\lim_{E\to0} \sigma^{(0)} (D,E) = 4e^2 / \pi h$, a
universal constant \cite{Titov:07}, in the presence of applied uniaxial strain
one finds
\begin{equation}
\lim_{E\to0} \sigma(D,E) = \frac{1}{\xi} \frac{4e^2}{\pi h}  .
\label{eq:sigmaEto0}
\end{equation}
Only in the case of hydrostatic strain ($\nu=-1$, $\lambda_x = \lambda_y$,
$\xi=1$) does one recover the universal limit, regardless of the strain modulus
\cite{Concha:10}. On the other hand, one finds $\lim_{E\to0} F(D,E) =
\frac{1}{3}$, corresponding to strongly sub-Poissonian noise
\cite{Tworzydlo:06}, regardless of applied strain.

In the opposite limit, the conductivity across a single barrier in the absence
of strain is linear in energy, $\sigma^{(0)} \approx (e^2 /h) D |E|/\hbar\vF$
for $E\to\infty$, with damped oscillations characterized by a pseudoperiod
$\Delta E$ such that \cite{Hannes:10} $D\Delta E/\hbar\vF = \pi$. In the
presence of strain, such results are modified by Eqs.~(\ref{eq:eff}), so that
$\sigma(E)\approx \sigma_\infty (E)$ for $E\to\infty$, with
\begin{equation}
\sigma_\infty (E) = \frac{4e^2}{h} \frac{D|E|}{4} \zeta ,
\label{eq:condasym}
\end{equation}
with damped oscillations characterized by a pseudoperiod given by
\begin{equation}
\xi \zeta D \frac{\Delta E}{\hbar\vF} = \pi.
\label{eq:condqperiod}
\end{equation}
In view of the fact that $|\lambda_x |> |\lambda_y |$, one may conclude that applied
strain induces a slight change in the slope of $\sigma$ \emph{vs} $|E|$, while
it modifies the pseudoperiod of the oscillations more substantially.

\begin{figure}[t]
\centering
\includegraphics[height=\columnwidth,angle=-90]{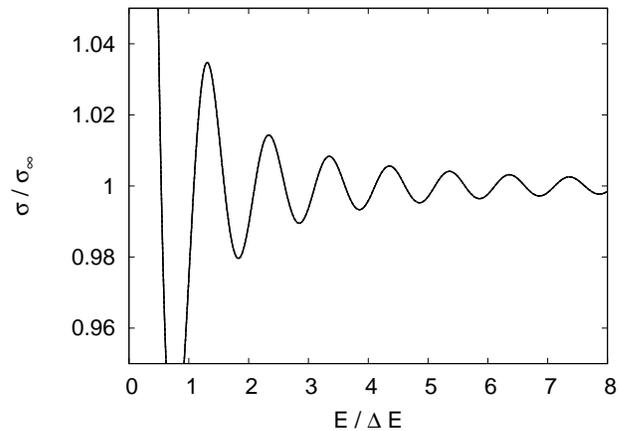}
\caption{Conductivity across a graphene strip ($D=100$~nm) normalized to
asymptotic large-energy behavior, Eq.~(\ref{eq:condasym}), \emph{vs.} energy
scaled to the pseudoperiod, Eq.~(\ref{eq:condqperiod}). Actually shown are four
curves, all collapsing into a single one, corresponding to strain applied along
the armchair direction ($\theta=\pi/2$), with $\varepsilon=0.03$, $0.05$,
$0.10$, $0.15$.}
\label{fig:singlecondarm}
\end{figure}

\begin{figure}[t]
\centering
\includegraphics[height=\columnwidth,angle=-90]{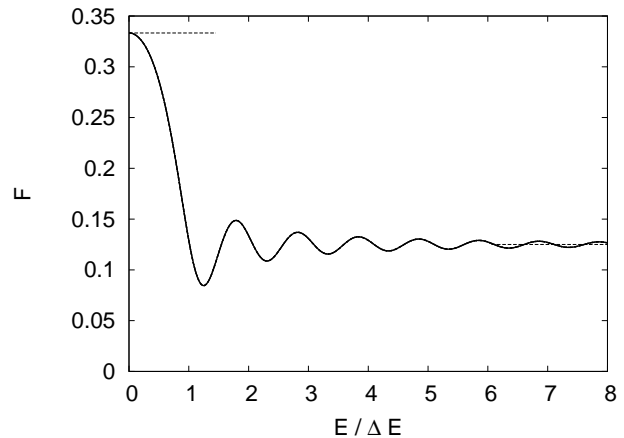}
\caption{Fano factor for ballistic transport across a graphene strip. All
parameters are as in Fig.~\ref{fig:singlecondarm}. Dashed lines represent the
universal low- and large-energy asymptotic values, $F(0)=\frac{1}{3}$ and $F_\infty
=\frac{1}{8}$, respectively.}
\label{fig:singlefanoarm}
\end{figure}

Fig.~\ref{fig:singlecondarm} shows our results for the scaled conductivity in
the presence of uniaxial strain ($\varepsilon = 0.03-0.15$) applied along the
armchair direction ($\theta=\pi/2$). When the conductivity $\sigma(E)$ is
normalized with respect to its asymptotic limit, Eq.~(\ref{eq:condasym}), and
plotted against energy $E$ scaled with the strain-dependent pseudoperiod $\Delta
E$, Eq.~(\ref{eq:condqperiod}), results corresponding to different values of the
strain modulus collapse into a single curve, displaying damped oscillations, as
prescribed by Eq.~(\ref{eq:condqperiod}). Similarly,
Fig.~\ref{fig:singlefanoarm} reports our results for the Fano factor as a
function of scaled energy. Again, the results for all the strain moduli here 
considered ($\varepsilon = 0.03-0.15$) collapse into a single, oscillating
curve. Note that the universal limits $F(E=0)= \frac{1}{3}$ and $F_\infty \equiv
\lim_{E\to\infty} F(E) = \frac{1}{8}$ are recovered in all cases, regardless of
applied strain. Such results do not depend on the direction $\theta$ of applied
strain.

\section{Transmission across multiple barriers}
\label{sec:multiple}

\begin{figure}[t]
\centering
\includegraphics[width=\columnwidth]{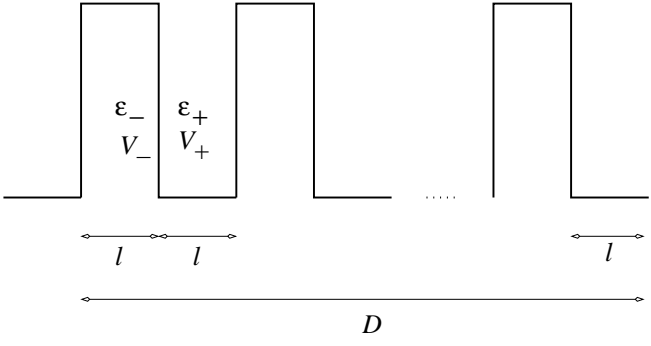}
\caption{Schematic plot of the multiple barrier, as considered in
Sec.~\ref{sec:multiple}.}
\label{fig:multibarrier}
\end{figure}

We next consider quasiparticle tunneling across $N$ identical barriers, each of
width $\ell$, two nearest neighbor (NN) barriers being separated by the distance
$\ell$, such that $2N\ell =D$ (Fig.~\ref{fig:multibarrier}). We assume a
position-dependent strain modulus $\varepsilon(x)$ and gate potential energy
$U(x)$, with 
\begin{subequations}
\begin{eqnarray}
\varepsilon(x) &=& \varepsilon_- , \quad (m-1)\ell\leq x\leq m\ell ,\\
               &=& \varepsilon_+ , \quad m\ell\leq x\leq (m+1)\ell ,
\end{eqnarray}
\end{subequations}
and 
\begin{subequations}
\begin{eqnarray}
U(x) &=& U_- , \quad (m-1)\ell\leq x\leq m\ell ,\\
     &=& U_+ , \quad m\ell\leq x\leq(m+1)\ell ,
\end{eqnarray}
\end{subequations}
with $m=1,\ldots 2N-1$. We
further consider the possibility of contacting the two extrema of the chain of
barriers with leads at the potential $V_L$. Eqs.~(\ref{eq:continuity}) then
suggest to look for a solution of the Dirac equation in the form
\begin{equation}
\psi(x,y) = U^\dag (\theta) \frac{\phi(x)}{\sqrt{1-\lambda_x \varepsilon(x)}}
e^{ik_y y}
\label{eq:Diracsolution}
\end{equation}
so that $\phi(x)$ is a continuous function at the barriers' edges. The
stationary Dirac equation for $\phi(x)$ can then be casted in the form of an
evolution equation \cite{Hannes:10}, so that $\phi(x)={\mathbb T}^{(N)} (x,x_0 )
\phi(x_0)$, where the evolution matrix ${\mathbb T}^{(N)} (x,x_0 )$ in turn obeys the
equation
\begin{eqnarray}
\frac{d}{dx} {\mathbb T}^{(N)} (x,x_0 ) &=& \left[
iq_{\mathrm{D}x}^{(0)} \varepsilon(x) \tau_z \mathbb{I} + \frac{i}{\hbar\vF}
\frac{E-U(x)}{1-\lambda_x \varepsilon(x)} \sigma_x \right.\nonumber\\
&&\hspace{-2.5truecm}\left.+ \frac{1-\lambda_y
\varepsilon(x)}{1-\lambda_x \varepsilon(x)} \left( k_y - q_{\mathrm{D}y}^{(0)}
\varepsilon(x)\tau_z \right) \sigma_z
\right] {\mathbb T}^{(N)} (x,x_0 ) ,
\label{eq:evolution}
\end{eqnarray}
with ${\mathbb T}^{(N)} (x_0 ,x_0 ) = {\mathbb I}$. For a single barrier, the
evolution matrix is related to the transfer matrix by \cite{Titov:07}
\begin{equation}
\mathbb{M}^{(1)} (x,x_0) = Q_s^{-1} (\varphi) \mathbb{T}^{(1)} (x,x_0 ) Q_s (\varphi),
\end{equation}
where
\begin{equation}
Q_s (\varphi) = \frac{1}{\sqrt{2}}\begin{pmatrix}  1 & 1 \\ se^{i\varphi} &
-se^{-i\varphi} \end{pmatrix}
\end{equation}
includes the incidence angle $\varphi$ of the incoming spinor,
Eq.~(\ref{eq:singlesols:I}), and $s=\sgn (E)$. In the limit of metallic leads
($|V_L |\to\infty$), one has $\varphi\to0$, with $Q_+ (0) = \frac{1}{\sqrt{2}}
(\sigma_z + \sigma_x)$, $Q_+^{-1} (0)  = Q_+ (0)$, and $Q_- (0)= Q_+
(0)\sigma_x$, $Q_-^{-1} (0) = \sigma_x Q_+ (0)$. The elements of the transfer
matrix can be furthermore related to the elements of the scattering matrix
across the barrier,
\begin{equation}
\mathbb{S} = \begin{pmatrix} r & t^\prime \\ t & r^\prime \end{pmatrix} ,
\end{equation}
where $r$, $t$ (\emph{resp.,} $r^\prime$, $t^\prime$) are the amplitudes of the
reflected and transmitted waves in region I (\emph{resp.,} III), cf.
Fig.~\ref{fig:barrier}. Indeed, one explicitly finds \cite{Titov:07,Bruus:04}
\begin{equation}
\mathbb{M}^{(1)}  = \begin{pmatrix} (t^\dag )^{-1} & r^\prime (t^\prime )^{-1} \\
-(t^\prime )^{-1} r & (t^\prime )^{-1} \end{pmatrix} .
\end{equation}
Therefore, for the conductance across a single barrier, one finds
\begin{equation}
G = \frac{2e^2}{h} \Tr (t^\dag t) = \frac{2e^2}{h} \Tr \left(
(\mathbb{M}^{(1)\dag}_{11}
\mathbb{M}^{(1)}_{11} )^{-1}
\right) ,
\label{eq:G}
\end{equation}
where $\Tr \equiv W \sum_\alpha \int_{-\infty}^\infty dk_y /2\pi$.
Correspondingly, the transmission for an incoming quasiparticle with energy $E$
and transverse wavevector $k_y$ in valley $\alpha$ is $T_\alpha (k_y) =
(\mathbb{M}^{(1)\dag}_{11} \mathbb{M}^{(1)}_{11} )^{-1}$, and the expressions
for the conductivity, Eq.~(\ref{eq:Landauer}), and Fano factor,
Eq.~(\ref{eq:Fano}), follow straightforwardly.

The solution of Eq.~(\ref{eq:evolution}) for the transfer matrix is derived
analytically in Appendix~\ref{app:transfer}, both for a single and for a
multiple barrier, in presence of strain-induced deformation of the Dirac cone.
Making use of Eqs.~(\ref{eq:app:T}) for the transmission $T_\alpha (k_y)$ in
Landauer's formula for the conductivity, Eq.~(\ref{eq:Landauer}), and in the
definition for the Fano factor, Eq.~(\ref{eq:Fano}), one again finds that the
conductivity in strained graphene, and strained graphene where the
strain-induced velocity anisotropy has been neglected, are related by means of
Eqs.~(\ref{eq:eff}), (\ref{eq:sigmaeff}), but now with $D=2N\ell$, and
\begin{subequations}
\begin{eqnarray}
\label{eq:ximulti}
\xi &=& \frac{1}{2} (\xi_+ + \xi_- ) ,\\
\zeta &=& \frac{1}{2} (\zeta_+ + \zeta_- ) ,\\
\xi_\pm &=& \frac{1- \lambda_y \varepsilon_\pm}{1-\lambda_x
\varepsilon_\pm} ,\\
\zeta_\pm &=& \frac{1}{1-\lambda_y \varepsilon_\pm} .
\end{eqnarray}
\end{subequations}
Eq.~(\ref{eq:sigmaEto0}) in the limit $E\to0$ then follows straightforwardly,
with $\xi$ given now by Eq.~(\ref{eq:ximulti}).
Moreover, the conductivity at large energies is characterized by an overall
linear behavior, interrupted by dips with decreasing depth, which result from the
coherent superposition of the damped oscillations produced by scattering off the
edges of the single barriers. The energies $E_n$ at which such dips occur are
asymptotically given by (cf. Appendix~\ref{app:transfer})
\begin{equation}
\frac{E_n}{\hbar\vF} \frac{D}{N} \frac{1}{2} ( \xi_+ \zeta_+ +  \xi_- \zeta_-) = n\pi,
\label{eq:dips}
\end{equation}
with $n$ an integer.

\begin{figure}[t]
\centering
\includegraphics[height=\columnwidth,angle=-90]{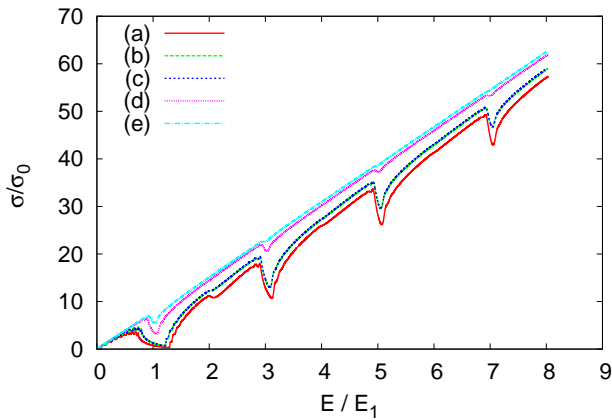}
\caption{(Color online) Conductivity $\sigma(E)$ in units of $\sigma_0 = 4 e^2
/h$, \emph{vs.} energy $E$, scaled with respect to the approximate location of
the first dip, $E_1$, as given by Eq.~(\ref{eq:dips}). Subsequent dips then
occur close to integer values of the ratio $E/E_1$. Uniaxial strain is
applied along the armchair direction ($\theta=\pi/2$) in the case of a
multibarrier superlattice, with $N=10$ barriers, $\ell=25$~nm ($D=500$~nm).
Different curves refer to nonuniform strain moduli within and outside NN
barriers (cf. Fig.~\ref{fig:multibarrier}), with 
(a) $\varepsilon_+ = 0.004$, $\varepsilon_- = 0$;
(b) $\varepsilon_+ = 0.003$, $\varepsilon_- = 0$;
(c) $\varepsilon_+ = 0.002$, $\varepsilon_- = -0.001$;
(d) $\varepsilon_+ = 0.002$, $\varepsilon_- =  0.001$;
(e) $\varepsilon_+ = 0.0005$, $\varepsilon_- = 0$.
In all cases, we set $U_\pm = 0$, for the sake of simplicity.}
\label{fig:multicondarm}
\end{figure}

\begin{figure}[t]
\centering
\includegraphics[height=\columnwidth,angle=-90]{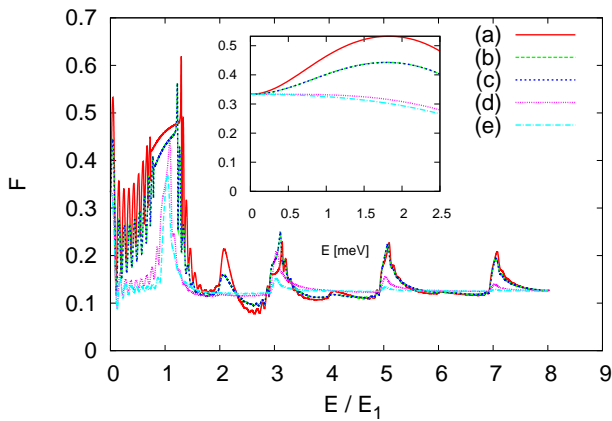}
\caption{(Color online) Fano factor $F$ \emph{vs.} scaled energy $E/E_1$, for
transport across a multibarrier superlattice, with nonuniform uniaxial strain
applied along the armchair direction ($\theta=\pi/2$). All parameters are as in
Fig.~\ref{fig:multicondarm}. Inset shows the universal low-energy asymptotic
behavior in the various cases. In the limit $E\to0$, the universal asymptotic
value, $F(0)=\frac{1}{3}$, is recovered.}
\label{fig:multifanoarm}
\end{figure}

\begin{figure}[t]
\centering
\includegraphics[height=\columnwidth,angle=-90]{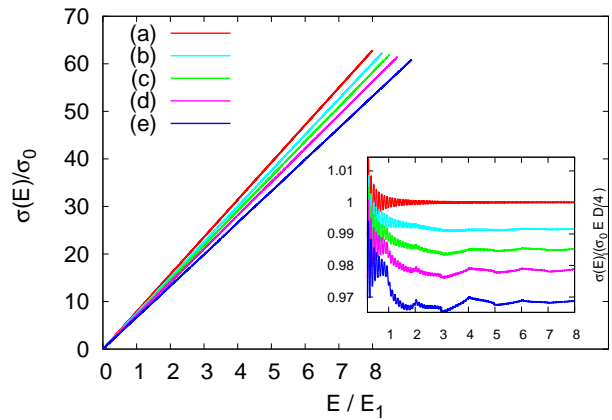}
\caption{(Color online) Conductivity $\sigma(E)$ in units of $\sigma_0 = 4 e^2
/h$, \emph{vs.} energy $E$, scaled with respect to $E_1$, as given by
Eq.~(\ref{eq:dips}). Uniaxial strain is applied along the zig~zag direction
($\theta=0$) in the case of a multibarrier superlattice, with $N=10$
barriers, $\ell=25$~nm ($D=500$~nm). Different curves refer to nonuniform strain
moduli within and outside NN barriers (cf. Fig.~\ref{fig:multibarrier}), with 
(a) $\varepsilon_+ = 0$, $\varepsilon_- = 0$;
(b) $\varepsilon_+ = 0.03$, $\varepsilon_- = 0$;
(c) $\varepsilon_+ = 0.05$, $\varepsilon_- = 0$;
(d) $\varepsilon_+ = 0.07$, $\varepsilon_- = 0$;
(e) $\varepsilon_+ = 0.10$, $\varepsilon_- = 0$.
In all cases, we set $U_\pm = 0$, for the sake of simplicity.
Inset shows the conductivity scaled with respect to its large-energy asymptotic
limit, $\sigma/\sigma_\infty$, as a function of scaled energy, $E/E_1$.
}
\label{fig:multicondzz}
\end{figure}

\begin{figure}[t]
\centering
\includegraphics[height=\columnwidth,angle=-90]{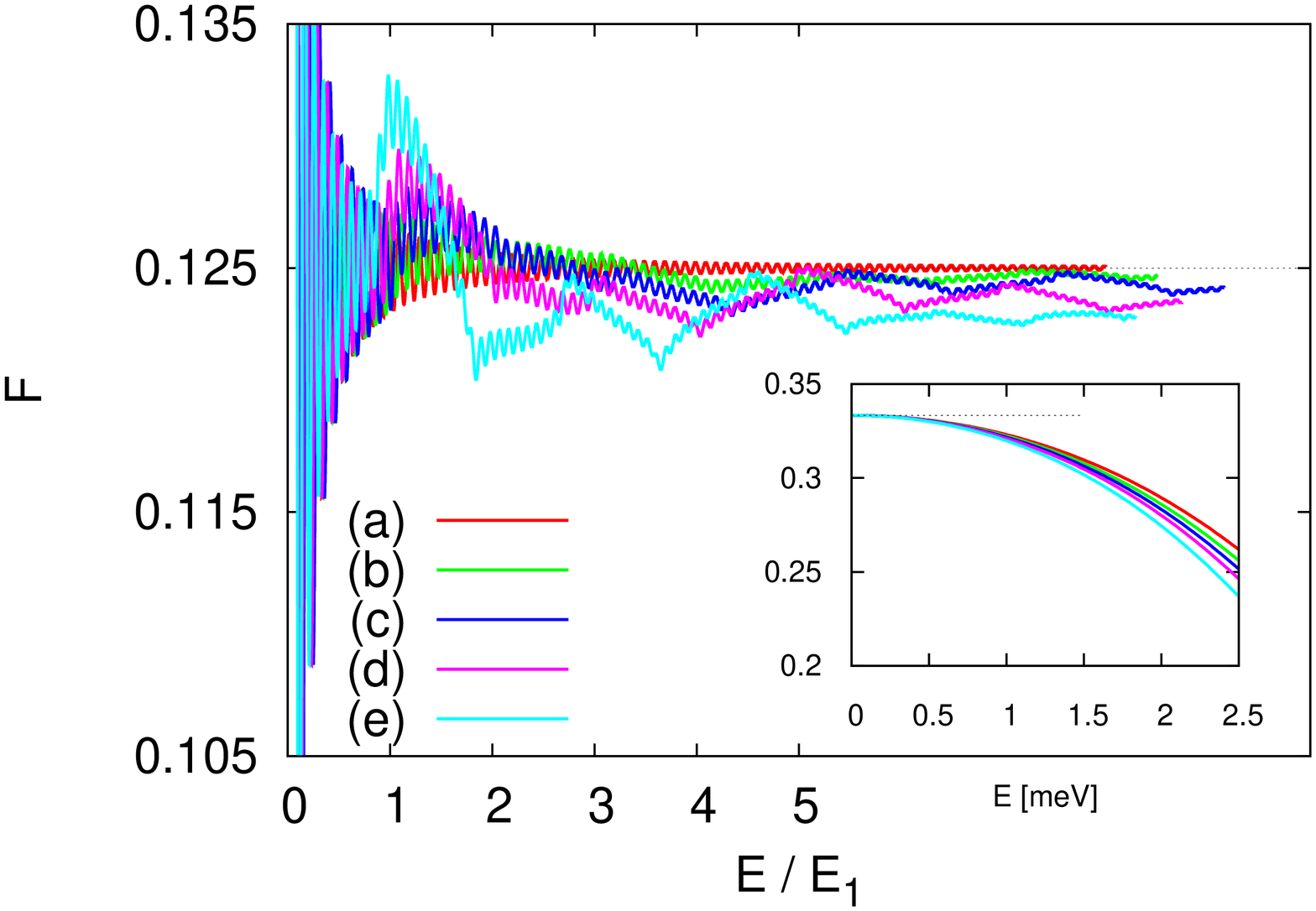}
\caption{(Color online) Fano factor $F$ \emph{vs.} scaled energy $E/E_1$, for
transport across a multibarrier superlattice, with nonuniform uniaxial strain
applied along the zig~zag direction ($\theta=0$). All parameters are as in
Fig.~\ref{fig:multicondzz}. Note the deviations from the large-energy asymptotic
limit for the unstrained case, $F_\infty = \frac{1}{8}$ (dashed line). The
low-energy universal limit, $F(0)=\frac{1}{3}$ (inset, dashed line), is
recovered, regardless of strain.}
\label{fig:multifanozz}
\end{figure}

Fig.~\ref{fig:multicondarm} shows our numerical results for the conductivity in
strained graphene, with strain applied nonuniformly along the armchair
direction, across a superlattice of $N=10$ barriers. At variance with
Fig.~\ref{fig:singlecondarm}, we have not scaled $\sigma$ with its asymptotic
behavior at large energies, Eq.~(\ref{eq:condasym}). As expected, the overall
linear behavior of $\sigma(E)$ is interrupted by dips, whose approximate energy
location is given by Eq.~(\ref{eq:dips}). While such dips get damped as energy
increases, they are nonetheless enhanced with respect to the case in which the
strain-induced deformation of the Dirac cones is neglected
\cite{Gattenloehner:10}, especially those corresponding to even integer values
of $n$ in Eq.~(\ref{eq:dips}). Correspondingly, the Fano factor
(Fig.~\ref{fig:multifanoarm}) is characterized by essentially analogous
features, with bumps occurring at approximately $E_n$, Eq.~(\ref{eq:dips}). In
particular, the universal limit at low energy, $F(0)=\frac{1}{3}$,
is recovered as in the
single-barrier case, regardless of applied strain.

Fig.~\ref{fig:multicondzz} shows our numerical results for the conductivity in
strained graphene, but now for nonuniform strain applied along the zig~zag
direction. At variance with the armchair case
(Fig.~\ref{fig:multicondarm}), for strain applied along the zig~zag direction
the conductivity seems not to be characterized by prominent dips as a function
of energy. This may explained by a reduced coherent superposition of the effects
due to each single barrier. However, if the trailing linear dependence on energy
is divided out (Fig.~\ref{fig:multicondzz}, inset), one may again recognize
`oscillations', with extrema approximatively occurring at $E_n$, as given by
Eq.~(\ref{eq:dips}). At variance with the armchair case, the Fano factor
exhibits a strain-dependent asymptotic limit, for large energies
(Fig.~\ref{fig:multifanozz}), with increasing deviations from the unstrained
behavior $F_\infty = \frac{1}{8}$, with increasing strain modulus $\varepsilon$
(at least within the strain range that has been numerically investigated). On
the other hand, both the oscillations as a function of scaled energy $E/E_1$ and
the low-energy limit $F(0) = \frac{1}{3}$ (Fig.~\ref{fig:multifanozz}, inset)
are recovered.

\section{Trasmission across a smooth barrier: effect of continuous strain}
\label{sec:smooth}

\begin{figure}[t]
\centering
\includegraphics[height=\columnwidth,angle=-90]{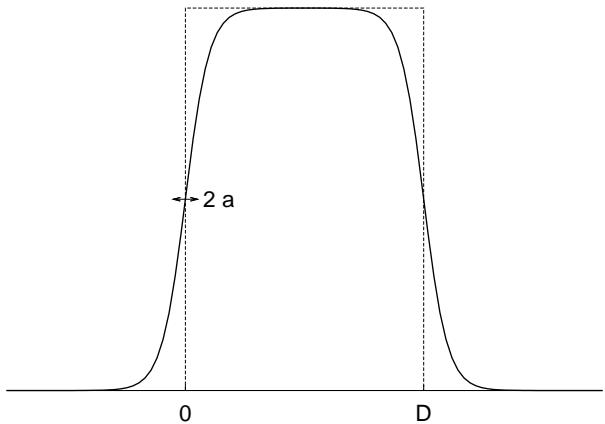}
\caption{Schematic single tunneling barrier, with smooth strain profile,
Eq.~(\ref{eq:smooth}). Dashed line depicts a sharp barrier, corresponding to the
limit $a\to0$.}
\label{fig:smooth}
\end{figure}

Although considerable insight is afforded by analytical solutions to the problem
of tunneling across single or multiple \emph{sharp} barriers, there is
sufficient evidence, both experimental \cite{Lee:08a} and theoretical
\cite{Cayssol:09}, that barrier edge effects are also important to determine the
transport properties across corrugated graphene. Here, we therefore consider the
case in which uniaxial strain is applied in a nonuniform but continuous fashion
to a graphene sheet, which can be modeled by a single barrier with \emph{smooth}
strain and gate potential profiles, $\varepsilon=\varepsilon(x)$ and $U=U(x)$,
respectively. Such a description includes and generalizes, in particular, a
continuous Fermi wavevector profile, as considered in
Ref.~\onlinecite{Cayssol:09}.

On quite general grounds, one may expect that a smooth potential profile
(whether induced by strain or by gating) introduces a new length scale, $a$ say
[as in Eq.~(\ref{eq:smooth}) below], which is the linear size over which the
potential strain varies appreciably. Such a new length scale has then to be
compared with the atomic scale, measured by the lattice step ${\frak a}$, on one
hand, and with the Fermi wavelength $\lambda_{\mathrm{F}} = \hbar\vF / (2\pi E)$
corresponding to the incident energy $E$, on the other. The approximation of a
sharp barrier (no smoothing) then holds  whenever $a\ll {\frak a} \ll
\lambda_{\mathrm{F}}$, \emph{i.e.} at sufficiently large incident energies. On
the other hand, the detailed structure of the barrier needs to be considered
when $a \sim \lambda_{\mathrm{F}}$. In both cases, we are interested to the more
general and realistic cases where $a\ll{\frak a}$, where one may additionally
neglect the occurrence of $\bK$--$\bK^\prime$ coupling. Indeed, truly sharp
electrostatic barriers on the order of the electron wavelength are quite
difficult to be realized, as is \emph{e.g.} demonstrated by the occurrence of
Fabry-P\'erot oscillations of the conductance in graphene heterostructures as
narrow as $\sim 20$~nm, where a resonant cavity is formed between two
electrostatically created bipolar junctions \cite{Young:09}. Such oscillations
are more accurately described when the smooth structure of these potential
barriers is taken into account, whereas intervalley scattering can be safely
neglected (see Supplementary Information in Ref.~\onlinecite{Young:09}). Another
instance of nonuniform barrier, where smoothing effects are important, is the
strain-induced ripples superlattice experimentally realized in
Ref.~\onlinecite{Bao:09}, which smoothing is essential on a length scale of
$\sim 100$~nm, whereas intervalley processes are negligible.

The kinetic part of the Hamiltonian for graphene subjected to uniform strain
$\varepsilon$ along the direction $\theta$ is
\begin{equation}
H = U^\dag (\theta) \sigma_i \hbar v_i \left( \frac{1}{i} \nabla_i
- q_{{\mathrm{D}}i} \right) U(\theta) ,
\label{eq:Hsmoothunstrained}
\end{equation}
where $v_i = \vF (1-\lambda_i \varepsilon)$, and summation over the repeated
index $i=1,2$ is understood. In order to generalize
Eq.~(\ref{eq:Hsmoothunstrained}) to the case of a nonuniform, but continuous
strain profile $\varepsilon=\varepsilon(x)$, one may be tempted to perform the
replacements $v_i \mapsto v_i (\br) \equiv \vF [1-\lambda_i \varepsilon(x)]$ and
$\bqD \mapsto \bqD (\br)$, Eq.~(\ref{eq:Diracrot}), with
$\varepsilon=\varepsilon(x)$. However, the resulting Hamiltonian must be
symmetrized, in order to preserve hermiticity, thus leading to the model
Hamiltonian for a nonuniform strain profile:
\begin{eqnarray}
H &=& U^\dag (\theta) \sigma_i \frac{1}{2} \left[ \hbar v_i (\br) \left( \frac{1}{i} \nabla_i -
q_{{\mathrm{D}}i} (\br) \right) \right.\nonumber\\
&&\left.+ \left( \frac{1}{i} \nabla_i -
q_{{\mathrm{D}}i} (\br) \right) \hbar v_i (\br) \right] U(\theta) .
\label{eq:Hsmoothstrain}
\end{eqnarray}
Eq.~(\ref{eq:Hsmoothstrain}) includes the effect of nonuniform, continuous
strain both as a shift in the position of the Dirac points, and as a deformation
of the Dirac cones (nonuniform and anisotropic Fermi velocity), at variance
\emph{e.g.} with Ref.~\onlinecite{Raoux:10}, where a nonuniform velocity is
considered, but an isotropic profile is assumed. As in the case of a single,
sharp barrier (Sec.~\ref{sec:single}), continuity of the current density,
Eqs.~(\ref{eq:continuity}), suggests to seek for a solution of the stationary
Dirac equation in a form analogous to Eq.~(\ref{eq:Diracsolution}), \emph{viz.}
\begin{equation}
\psi(x,y) = U^\dag (\theta) \frac{\phi(x)}{\sqrt{v_x (x)}}
e^{ik_y y} .
\end{equation}
One explicitly finds [cf. Eq.~(\ref{eq:evolution})]
\begin{eqnarray}
\frac{d\phi(x)}{dx} &=& \left[\frac{1-\lambda_y \varepsilon(x)}{1-\lambda_x
\varepsilon(x)} \left(k_y - q_{{\mathrm{D}}y}^{(0)} \varepsilon(x)\right) \sigma_z 
\right.\nonumber\\
&&\hspace{-1truecm}\left.
+ i \frac{E-U(x)}{\left(1-\lambda_x \varepsilon(x)\right)\hbar\vF} \sigma_x
+ i q_{{\mathrm{D}}x}^{(0)} \varepsilon(x) \mathbb{I} \right] \phi(x) .
\label{eq:evolution-phi}
\end{eqnarray}

We have solved Eq.~(\ref{eq:evolution-phi}) numerically, for the nonuniform,
smooth strain profile 
\begin{equation}
\varepsilon(x) = \frac{\varepsilon_0}{\tanh(D/4a)} \left( \frac{1}{1+e^{-x/a}} -
\frac{1}{1+e^{-(x-D)/a}} \right) ,
\label{eq:smooth}
\end{equation}
as shown in Fig.~\ref{fig:smooth}. Such a strain profile is essentially flat for
$|x-D/2|\ll a$, where $\varepsilon(x)\approx\varepsilon_0$, and for $|x-D/2|\gg
a$, where $\varepsilon(x)\approx 0$. In the limit $a/D\to0$,
Eq.~(\ref{eq:smooth}) tends to the sharp barrier considered in
Sec.~\ref{sec:single}. Therefore, asymptotically for $|x|\to\infty$, the
solutions of Eq.~(\ref{eq:evolution-phi}) must merge into
Eqs.~(\ref{eq:singlesols}), in regions I and III. We have therefore taken an
initial value $\phi(x=x_0 )$ in the form of Eq.~(\ref{eq:singlesols:III}), for
$x_0 = 5 D$, and integrated Eq.~(\ref{eq:evolution-phi}) backwards for $x\ll 0$.
Comparing the numerical solution with Eq.~(\ref{eq:singlesols:I}), one may
extract the reflection coefficient $r$, relative to an incident wave with unit
amplitude incoming from $x>0$, as the Fourier weight with respect to its
negative frequency component, whence the transmission $T(\phi)$ follows
straightforwardly. As a cross-check of our procedure, we have also verified that
the continuity equation, Eq.~(\ref{eq:continuity}), holds true, within the
numerical error.

\begin{figure}[t]
\centering
\includegraphics[height=\columnwidth,angle=-90]{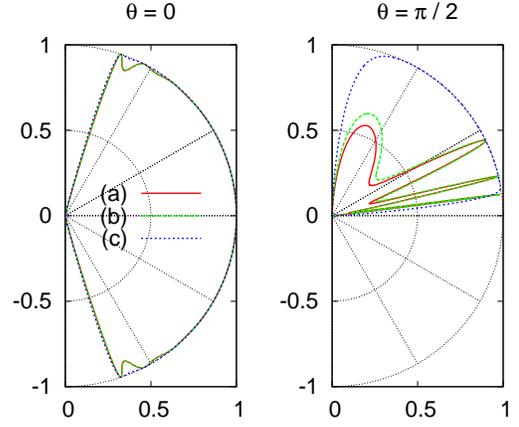}
\caption{(Color online) Tunneling transmission \emph{vs} incidence angle
$\varphi$ across a smooth strain barrier, Eq.~(\ref{eq:smooth}), with
$D=100$~nm, and incidence energy $E=80$~meV ($\lambda_{\mathrm{F}} = \hbar\vF /
(2\pi E) \approx 1.3$~nm). Left panel refers to strain applied along the zig~zag
direction ($\theta=0$), with $\varepsilon_0 = 0.1$. Right panel refers to strain
applied along the armchair direction ($\theta=0$), with $\varepsilon_0 = 0.01$.
In both cases, the different lines correspond to different values of the
smoothing parameter, \emph{viz.} (a) $a=0$ (sharp barrier); (b) $a=10^{-2} D =
1$~nm; (c) $a=10^{-1} D = 10$~nm. In all cases, $U(x)=0$, for the sake of
simplicity.}
\label{fig:singlesmooth80}
\end{figure}

\begin{figure}[t]
\centering
\includegraphics[height=\columnwidth,angle=-90]{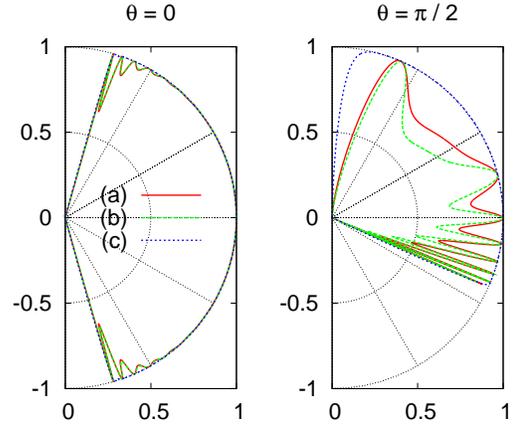}
\caption{(Color online) Same as Fig.~\ref{fig:singlesmooth80}, but with
$E=150$~meV ($\lambda_{\mathrm{F}} \approx 0.7$~nm).}
\label{fig:singlesmooth150}
\end{figure}

Figs.~\ref{fig:singlesmooth80} and \ref{fig:singlesmooth150} show our numerical
results for the tunneling transmission $T(\varphi)$ across the smooth strain
barrier, Eq.~(\ref{eq:smooth}), with $D=100$~nm and different values of the
smoothing parameter, $a/D$. Fig.~\ref{fig:singlesmooth80} refers to an incidence
energy $E=80$~meV, corresponding to an incident wavelength $\lambda_{\mathrm{F}}
= \hbar\vF / (2\pi E) \approx 1.3$~nm. One finds that transmission of
propagating waves is allowed for incidence angles $\varphi$ such that 
$\varphi_{\mathrm{cr}-} \leq \varphi \leq \varphi_{\mathrm{cr}+}$, with
\begin{equation}
\varphi_{\mathrm{cr}\pm} = \pm \arcsin \left( \frac{1}{1-\lambda_y
\varepsilon_0} \right) ,
\label{eq:phicritzz}
\end{equation}
in the zig~zag case ($\theta=0$), and
$\varphi >
\varphi_{\mathrm{cr}}$, with
\begin{equation}
\arcsin \left( - \frac{1}{1-\lambda_y \varepsilon_0} + \frac{\hbar\vF}{|E|}
\varepsilon_0 \kappa (1-\nu) \right) ,
\end{equation}
in the armchair case ($\theta=\pi/2$), independent of the smoothing parameter
$a/D$. Outside that window, transmission takes place via evanescent waves only,
and $T(\varphi)\approx 0$. For strain applied along the zig~zag direction
($\theta=0$, Fig.~\ref{fig:singlesmooth80}, left panel),
Eq.~(\ref{eq:phicritzz}) predicts the existence of critical angles
$|\varphi_{\mathrm{cr}\pm} | < \pi/2$. This is a direct consequence of the
strain-induced deformation of the Dirac cones [$\lambda_y \neq 0$ in
Eq.~(\ref{eq:phicritzz})].  Both in case of strain applied along the zig~zag and
armchair directions, increasing the smoothness parameter $a/D$ away from the
limit of a sharp barrier ($a/D=0$) suppresses the oscillations in $T(\varphi)$
within the propagating window, until $a > \lambda_{\mathrm{F}}$, in which
case transmission is almost undisturbed by the presence of the barrier. These
results are confirmed by Fig.~\ref{fig:singlesmooth150}, where we consider
quasiparticles with larger incident energy $E=150$~meV, corresponding to a
smaller Fermi wavelength $\lambda_{\mathrm{F}} \approx 0.7$~nm. While the
transmission window widens and the number of oscillations increases, smoothening
the strain profile immediately washes out the deviations of the tunneling
transmission from unity. In ending this section, we note that the procedure
applied to extracting the tunneling transmission from the numerical solution of
Eq.~(\ref{eq:evolution-phi}) can be generalized, in principle, to the case of an
arbitrary nonuniform strain potential, such as a superlattice of several smooth
barriers, such as Eq.~(\ref{eq:smooth}).

\section{Conclusions}
\label{sec:conclusions}

We have studied the effect of a strain-induced modulation of the Fermi velocity
on several transport properties of graphene, such as the angular dependence of
the tunneling transmission, the conductivity, and the Fano factor. After
considering the cases of a single sharp tunneling barrier, and of a
superstructure of several, periodically repeated, such sharp barriers, we have
specifically studied the case in which both the modulus of applied uniaxial
strain, and possibly an applied gate potential, depend continuously on position.
This is expected to afford a more accurate description of real `origami' device
\cite{Pereira:09}, in which `foldings' of a graphene sheet would conceivably
involve several lattice spacings. In the case of sharp tunneling barriers, we
have demonstrated that the effect of a strain-induced deformation of the Dirac
cone is of the same order of the strain-induced shift of the Dirac points, and
should therefore be taken into account on the same basis. In particular, we have
found that strain modifies the quasi-period in energy that regulates the
occurrence of dips in the conductivity across a superstructure of several sharp
barriers, due to coherent scattering off their edges. Such effect is however
less dramatic in the energy dependence of the Fano factor. Finally, we have
generalized our results to embrace the case of a generic \emph{nonuniform}
strain, and possibly a gate potential, profile. Besides allowing a more accurate
analysis of tunneling transmission across smooth barriers, especially at low
incident energies, which are expected to be more sensitive to local deviations
from uniformity, such an approach can be applied to describe arbitrary strain
superstructures, albeit numerically.

Among the already available experimental results which could be described in
terms of a strain-induced deformation of the Dirac cones, we mention Raman
spectroscopy \cite{Huang:10} and the strain-dependence of both the longitudinal
and the recently predicted transverse plasmon mode
\cite{Mikhailov:07,Pellegrino:11}. Moreover, transmittance measurements with
polarized light between the near-infrared and the ultraviolet on uniaxially
strained graphene may provide information on the Dirac cone deformation
\cite{Nair:08,Pellegrino:11}.

\acknowledgments

FMDP acknowledges Dr D. M. Basko for discussions and correspondence over the
general area embraced by the present work.

\appendix
\section{Transfer matrix across a multiple barrier}
\label{app:transfer}

In the case of a single barrier ($N=1$, $2\ell = D$),
Fig.~\ref{fig:multibarrier}), Eq.~(\ref{eq:evolution}) for the transfer matrix
admits the analytical solution
\begin{eqnarray}
\mathbb{M}^{(1)} (D,0) &=& \exp \left( i q_{\mathrm{D}x}^{(0)} \varepsilon D
\right)
\exp \left(
\frac{i}{\hbar\vF} \frac{(E-U_g)D}{1-\lambda_x \varepsilon}\sigma_z
\right.\nonumber\\
&&\left.
+ \frac{1-\lambda_y \varepsilon}{1-\lambda_x \varepsilon} (k_y -
q_{\mathrm{D}y}^{(0)} )D \sigma_ x \right) ,
\label{eq:app:single}
\end{eqnarray}
corresponding to the initial condition $\mathbb{M}^{(1)} (0,0)=\mathbb{I}$, and
to a uniform strain $\varepsilon$ and to a gate potential energy $U_g$ across
the barrier. The second matrix exponential in Eq.~(\ref{eq:app:single}) can be
made more explicit, by making use of the following identity for a linear
combination of the Pauli matrices,
\begin{eqnarray}
\exp( \mathbf{a}\cdot\bsigma ) &=& \frac{\sinh a}{a} \mathbf{a}\cdot\bsigma +
\mathbb{I} \cosh a ,
\end{eqnarray}
where $a=(\sum_i a_i^2 )^{1/2}$, and $a_i \in\mathbb{C}$ ($i=1,2,3$).

We next consider a single barrier, but now with nonuniform strain modulus and
gate potential energy, \emph{i.e.} $\varepsilon(x) = \varepsilon_-$ and
$U(x)=U_-$ within the barrier ($0< x < \ell$), and $\varepsilon(x) =
\varepsilon_+$ and $U(x)=U_+$ beyond the barrier's second edge ($\ell<x<2\ell$;
cf. Fig.~\ref{fig:multibarrier}). In this case, one finds $\mathbb{M}^{(1)}
(2\ell,0) = \mathbb{M}_+ (\ell) \mathbb{M}_- (\ell)$, where $\mathbb{M}_\pm
(\ell)$ are given by Eq.~(\ref{eq:app:single}), with $D\mapsto\ell$,
$\varepsilon\mapsto\varepsilon_\pm$, and $U\mapsto U_\pm$. One finds
\begin{equation}
\mathbb{M}^{(1)} (2\ell,0) = 
e^{i q_{\mathrm{D}x}^{(0)} (\varepsilon_+ + \varepsilon_- )\ell}
\tilde{\mathbb{M}}_1 ,
\label{eq:unimodular}
\end{equation}
where $\tilde{\mathbb{M}}_1$ is a unimodular matrix, $\det
\tilde{\mathbb{M}}_1 = 1$. Specifically, one finds
\begin{equation}
\left(\tilde{\mathbb{M}}_1 \right)_{11} = \lambda+i\eta
,
\end{equation}
where
\begin{subequations}
\begin{eqnarray}
\lambda &=&
\frac{\sinh (q_- \ell)}{q_-}
\frac{\sinh (q_+ \ell)}{q_+}
(\kappa_- \kappa_+ - u_- u_+ ) \nonumber\\
&&+
\cosh (q_- \ell) \cosh (q_+ \ell) ,\\
\eta &=&
\frac{u_-}{q_-} \sinh (q_- \ell ) \cosh (q_+ \ell) \nonumber\\
&&+
\frac{u_+}{q_+} \sinh (q_+ \ell ) \cosh (q_- \ell),
\end{eqnarray}
\end{subequations}
with
\begin{subequations}
\begin{eqnarray}
\kappa_\pm &=& \frac{1 - \lambda_y \varepsilon_\pm}{1 - 
\lambda_x \varepsilon_\pm} (k_y - q_{\mathrm{D}y}^{(0)} \varepsilon_\pm ) ,\\
u_\pm &=& \frac{E-U_\pm}{\hbar\vF (1 - \lambda_x \varepsilon_\pm )} , \\
q_\pm &=& \sqrt{\kappa_\pm^2 - u_\pm^2} ,
\end{eqnarray}
\end{subequations}
whence Eq.~(\ref{eq:G}) follows straightforwardly.

Finally, in the case of $N$ barriers ($D=2N\ell$, Fig.~\ref{fig:multibarrier}),
iterating Eq.~(\ref{eq:unimodular}) $N$ times, one has
\begin{equation}
\mathbb{M}^{(N)} (D,0) = 
e^{i q_{\mathrm{D}x}^{(0)} (\varepsilon_+ + \varepsilon_- )N\ell}
\tilde{\mathbb{M}}_1^N ,
\end{equation}
where for the $N$th power of the unimodular matrix $\tilde{\mathbb{M}}_1$ one
may use an identity due to Chebyshev \cite{Yeh:77}, and specifically obtain
\begin{equation}
\left(\tilde{\mathbb{M}}_1^N \right)_{11} =
\frac{\sinh (Nz)}{\sinh z} ( \tilde{\mathbb{M}}_1 )_{11}
-
\frac{\sinh((N-1)z)}{\sinh z} .
\end{equation}
Here, we have denoted the eigenvalues of $\tilde{\mathbb{M}}_1$ by $e^{\pm
z}$, with $z\in\mathbb{C}$. Finally, one finds for the transmission 
\begin{eqnarray}
T(k_y) &=& 
\left[\left(\mathbb{M}^{(N)} (D,0) \right)_{11}^\ast
\left(\mathbb{M}^{(N)} (D,0) \right)_{11} \right]^{-1} \nonumber\\
&=& \left[ \cosh^2 (Nz) + \frac{\eta^2}{\lambda^2 -1} \sinh^2 (Nz) \right]^{-1}
.
\end{eqnarray}
Since $\lambda = \frac{1}{2} \Tr \tilde{\mathbb{M}}_1 = \cosh z$, one finds
explicitly
\begin{subequations}
\label{eq:app:T}
\begin{align}
T_\alpha (k_y ) &\equiv T^{\mathrm{prop}}_\alpha (k_y ) \nonumber\\
&= \left[\cos^2 (Ny) + \frac{\eta^2}{\lambda^2 -1} \sin^2 (Ny) \right]^{-1} ,\\
\intertext{with $y=\arccos \lambda$, if $|\lambda|<1$,} 
        &\equiv T^{\mathrm{evan}}_\alpha (k_y ) \nonumber\\
&= \left[\cosh^2 (Nx) + \frac{\eta^2}{\lambda^2 -1} \sinh^2 (Nx) \right]^{-1} ,\\
\intertext{with $x = \log|\lambda+\sqrt{\lambda^2 -1}|$, if $|\lambda|>1$,}
        &= [1 + \eta^2 N^2 ]^{-1},
\end{align}
\end{subequations}
if $|\lambda|=1$. In particular, one finds $\lambda \sim \cos (u_+ \ell + u_-
\ell )$, for $E\to\infty$, whence Eq.~(\ref{eq:dips}) follows.

\begin{small} 
\bibliographystyle{apsrev}
\bibliography{a,b,c,d,e,f,g,h,i,j,k,l,m,n,o,p,q,r,s,t,u,v,w,x,y,z,zzproceedings,Angilella}

\begin{thebibliography}{58}
\expandafter\ifx\csname natexlab\endcsname\relax\def\natexlab#1{#1}\fi
\expandafter\ifx\csname bibnamefont\endcsname\relax
  \def\bibnamefont#1{#1}\fi
\expandafter\ifx\csname bibfnamefont\endcsname\relax
  \def\bibfnamefont#1{#1}\fi
\expandafter\ifx\csname citenamefont\endcsname\relax
  \def\citenamefont#1{#1}\fi
\expandafter\ifx\csname url\endcsname\relax
  \def\url#1{\texttt{#1}}\fi
\expandafter\ifx\csname urlprefix\endcsname\relax\def\urlprefix{URL }\fi
\providecommand{\bibinfo}[2]{#2}
\providecommand{\eprint}[2][]{\url{#2}}

\bibitem[{\citenamefont{Novoselov et~al.}(2005)\citenamefont{Novoselov, Jiang,
  Schedin, Booth, Khotkevich, Morozov, and Geim}}]{Novoselov:05a}
\bibinfo{author}{\bibfnamefont{K.~S.} \bibnamefont{Novoselov}},
  \bibinfo{author}{\bibfnamefont{D.}~\bibnamefont{Jiang}},
  \bibinfo{author}{\bibfnamefont{F.}~\bibnamefont{Schedin}},
  \bibinfo{author}{\bibfnamefont{T.~J.} \bibnamefont{Booth}},
  \bibinfo{author}{\bibfnamefont{V.~V.} \bibnamefont{Khotkevich}},
  \bibinfo{author}{\bibfnamefont{S.~V.} \bibnamefont{Morozov}},
  \bibnamefont{and} \bibinfo{author}{\bibfnamefont{A.~K.} \bibnamefont{Geim}},
  \bibinfo{journal}{Proc. Nat. Acad. Sci.} \textbf{\bibinfo{volume}{102}},
  \bibinfo{pages}{10451} (\bibinfo{year}{2005}).

\bibitem[{\citenamefont{{Castro Neto} et~al.}(2009)\citenamefont{{Castro Neto},
  Guinea, Peres, Novoselov, and Geim}}]{CastroNeto:08}
\bibinfo{author}{\bibfnamefont{A.~H.} \bibnamefont{{Castro Neto}}},
  \bibinfo{author}{\bibfnamefont{F.}~\bibnamefont{Guinea}},
  \bibinfo{author}{\bibfnamefont{N.~M.~R.} \bibnamefont{Peres}},
  \bibinfo{author}{\bibfnamefont{K.~S.} \bibnamefont{Novoselov}},
  \bibnamefont{and} \bibinfo{author}{\bibfnamefont{A.~K.} \bibnamefont{Geim}},
  \bibinfo{journal}{Rev. Mod. Phys.} \textbf{\bibinfo{volume}{81}},
  \bibinfo{pages}{000109} (\bibinfo{year}{2009}).

\bibitem[{\citenamefont{Abergel et~al.}(2010)\citenamefont{Abergel, Apalkov,
  Berashevich, Ziegler, and Chakraborty}}]{Abergel:10}
\bibinfo{author}{\bibfnamefont{D.~S.~L.} \bibnamefont{Abergel}},
  \bibinfo{author}{\bibfnamefont{V.}~\bibnamefont{Apalkov}},
  \bibinfo{author}{\bibfnamefont{J.}~\bibnamefont{Berashevich}},
  \bibinfo{author}{\bibfnamefont{K.}~\bibnamefont{Ziegler}}, \bibnamefont{and}
  \bibinfo{author}{\bibfnamefont{T.}~\bibnamefont{Chakraborty}},
  \bibinfo{journal}{Adv. Phys.} \textbf{\bibinfo{volume}{59}},
  \bibinfo{pages}{261} (\bibinfo{year}{2010}).

\bibitem[{\citenamefont{Pereira et~al.}(2006)\citenamefont{Pereira, Mlinar,
  Peeters, and Vasilopoulos}}]{MiltonPereira:06}
\bibinfo{author}{\bibfnamefont{J.~M.} \bibnamefont{Pereira}},
  \bibinfo{author}{\bibfnamefont{V.}~\bibnamefont{Mlinar}},
  \bibinfo{author}{\bibfnamefont{F.~M.} \bibnamefont{Peeters}},
  \bibnamefont{and}
  \bibinfo{author}{\bibfnamefont{P.}~\bibnamefont{Vasilopoulos}},
  \bibinfo{journal}{Phys. Rev. B} \textbf{\bibinfo{volume}{74}},
  \bibinfo{pages}{045424} (\bibinfo{year}{2006}).

\bibitem[{\citenamefont{Barbier et~al.}(2009)\citenamefont{Barbier,
  Vasilopoulos, and Peeters}}]{Barbier:09}
\bibinfo{author}{\bibfnamefont{M.}~\bibnamefont{Barbier}},
  \bibinfo{author}{\bibfnamefont{P.}~\bibnamefont{Vasilopoulos}},
  \bibnamefont{and} \bibinfo{author}{\bibfnamefont{F.~M.}
  \bibnamefont{Peeters}}, \bibinfo{journal}{Phys. Rev. B}
  \textbf{\bibinfo{volume}{80}}, \bibinfo{pages}{205415}
  (\bibinfo{year}{2009}).

\bibitem[{\citenamefont{Barbier
  et~al.}(2010{\natexlab{a}})\citenamefont{Barbier, Vasilopoulos, and
  Peeters}}]{Barbier:10}
\bibinfo{author}{\bibfnamefont{M.}~\bibnamefont{Barbier}},
  \bibinfo{author}{\bibfnamefont{P.}~\bibnamefont{Vasilopoulos}},
  \bibnamefont{and} \bibinfo{author}{\bibfnamefont{F.~M.}
  \bibnamefont{Peeters}}, \bibinfo{journal}{Phys. Rev. B}
  \textbf{\bibinfo{volume}{81}}, \bibinfo{pages}{075438}
  (\bibinfo{year}{2010}{\natexlab{a}}).

\bibitem[{\citenamefont{Barbier
  et~al.}(2010{\natexlab{b}})\citenamefont{Barbier, Vasilopoulos, and
  Peeters}}]{Barbier:10a}
\bibinfo{author}{\bibfnamefont{M.}~\bibnamefont{Barbier}},
  \bibinfo{author}{\bibfnamefont{P.}~\bibnamefont{Vasilopoulos}},
  \bibnamefont{and} \bibinfo{author}{\bibfnamefont{F.~M.}
  \bibnamefont{Peeters}}, \bibinfo{journal}{Phys. Rev. B}
  \textbf{\bibinfo{volume}{82}}, \bibinfo{pages}{235408}
  (\bibinfo{year}{2010}{\natexlab{b}}).

\bibitem[{\citenamefont{Peres}(2009)}]{Peres:09}
\bibinfo{author}{\bibfnamefont{N.~M.~R.} \bibnamefont{Peres}},
  \bibinfo{journal}{J. Phys.: Cond. Matter} \textbf{\bibinfo{volume}{21}},
  \bibinfo{pages}{323201} (\bibinfo{year}{2009}).

\bibitem[{\citenamefont{Nair et~al.}(2008)\citenamefont{Nair, Blake,
  Grigorenko, Novoselov, Booth, Stauber, Peres, and Geim}}]{Nair:08}
\bibinfo{author}{\bibfnamefont{R.~R.} \bibnamefont{Nair}},
  \bibinfo{author}{\bibfnamefont{P.}~\bibnamefont{Blake}},
  \bibinfo{author}{\bibfnamefont{A.~N.} \bibnamefont{Grigorenko}},
  \bibinfo{author}{\bibfnamefont{K.~S.} \bibnamefont{Novoselov}},
  \bibinfo{author}{\bibfnamefont{T.~J.} \bibnamefont{Booth}},
  \bibinfo{author}{\bibfnamefont{T.}~\bibnamefont{Stauber}},
  \bibinfo{author}{\bibfnamefont{N.~M.~R.} \bibnamefont{Peres}},
  \bibnamefont{and} \bibinfo{author}{\bibfnamefont{A.~K.} \bibnamefont{Geim}},
  \bibinfo{journal}{Science} \textbf{\bibinfo{volume}{320}},
  \bibinfo{pages}{1308} (\bibinfo{year}{2008}).

\bibitem[{\citenamefont{Kuzmenko et~al.}(2008)\citenamefont{Kuzmenko, van
  Heumen, Carbone, and van~der Marel}}]{Kuzmenko:08}
\bibinfo{author}{\bibfnamefont{A.~B.} \bibnamefont{Kuzmenko}},
  \bibinfo{author}{\bibfnamefont{E.}~\bibnamefont{van Heumen}},
  \bibinfo{author}{\bibfnamefont{F.}~\bibnamefont{Carbone}}, \bibnamefont{and}
  \bibinfo{author}{\bibfnamefont{D.}~\bibnamefont{van~der Marel}},
  \bibinfo{journal}{Phys. Rev. Lett.} \textbf{\bibinfo{volume}{100}},
  \bibinfo{pages}{117401} (\bibinfo{year}{2008}).

\bibitem[{\citenamefont{Wang et~al.}(2008)\citenamefont{Wang, Zhang, Tian,
  Girit, Zettl, Crommie, and Shen}}]{Wang:08}
\bibinfo{author}{\bibfnamefont{F.}~\bibnamefont{Wang}},
  \bibinfo{author}{\bibfnamefont{Y.}~\bibnamefont{Zhang}},
  \bibinfo{author}{\bibfnamefont{C.}~\bibnamefont{Tian}},
  \bibinfo{author}{\bibfnamefont{C.}~\bibnamefont{Girit}},
  \bibinfo{author}{\bibfnamefont{A.}~\bibnamefont{Zettl}},
  \bibinfo{author}{\bibfnamefont{M.}~\bibnamefont{Crommie}}, \bibnamefont{and}
  \bibinfo{author}{\bibfnamefont{Y.~R.} \bibnamefont{Shen}},
  \bibinfo{journal}{Science} \textbf{\bibinfo{volume}{320}},
  \bibinfo{pages}{206} (\bibinfo{year}{2008}).

\bibitem[{\citenamefont{Mak et~al.}(2008)\citenamefont{Mak, Sfeir, Wu, Lui,
  Misewich, and Heinz}}]{Mak:08}
\bibinfo{author}{\bibfnamefont{K.~F.} \bibnamefont{Mak}},
  \bibinfo{author}{\bibfnamefont{M.~Y.} \bibnamefont{Sfeir}},
  \bibinfo{author}{\bibfnamefont{Y.}~\bibnamefont{Wu}},
  \bibinfo{author}{\bibfnamefont{C.~H.} \bibnamefont{Lui}},
  \bibinfo{author}{\bibfnamefont{J.~A.} \bibnamefont{Misewich}},
  \bibnamefont{and} \bibinfo{author}{\bibfnamefont{T.~F.} \bibnamefont{Heinz}},
  \bibinfo{journal}{Phys. Rev. Lett.} \textbf{\bibinfo{volume}{101}},
  \bibinfo{pages}{196405} (\bibinfo{year}{2008}).

\bibitem[{\citenamefont{Stauber et~al.}(2008)\citenamefont{Stauber, Peres, and
  Geim}}]{Stauber:08a}
\bibinfo{author}{\bibfnamefont{T.}~\bibnamefont{Stauber}},
  \bibinfo{author}{\bibfnamefont{N.~M.~R.} \bibnamefont{Peres}},
  \bibnamefont{and} \bibinfo{author}{\bibfnamefont{A.~K.} \bibnamefont{Geim}},
  \bibinfo{journal}{Phys. Rev. B} \textbf{\bibinfo{volume}{78}},
  \bibinfo{pages}{085432} (\bibinfo{year}{2008}).

\bibitem[{\citenamefont{Pellegrino
  et~al.}(2010{\natexlab{a}})\citenamefont{Pellegrino, Angilella, and
  Pucci}}]{Pellegrino:09b}
\bibinfo{author}{\bibfnamefont{F.~M.~D.} \bibnamefont{Pellegrino}},
  \bibinfo{author}{\bibfnamefont{G.~G.~N.} \bibnamefont{Angilella}},
  \bibnamefont{and} \bibinfo{author}{\bibfnamefont{R.}~\bibnamefont{Pucci}},
  \bibinfo{journal}{Phys. Rev. B} \textbf{\bibinfo{volume}{81}},
  \bibinfo{pages}{035411} (\bibinfo{year}{2010}{\natexlab{a}}).

\bibitem[{\citenamefont{Hwang and Das~Sarma}(2007)}]{Hwang:07a}
\bibinfo{author}{\bibfnamefont{E.~H.} \bibnamefont{Hwang}} \bibnamefont{and}
  \bibinfo{author}{\bibfnamefont{S.}~\bibnamefont{Das~Sarma}},
  \bibinfo{journal}{Phys. Rev. B} \textbf{\bibinfo{volume}{75}},
  \bibinfo{pages}{205418} (\bibinfo{year}{2007}).

\bibitem[{\citenamefont{Polini et~al.}(2009)\citenamefont{Polini, {MacDonald},
  and Vignale}}]{Polini:09}
\bibinfo{author}{\bibfnamefont{M.}~\bibnamefont{Polini}},
  \bibinfo{author}{\bibfnamefont{A.~H.} \bibnamefont{{MacDonald}}},
  \bibnamefont{and} \bibinfo{author}{\bibfnamefont{G.}~\bibnamefont{Vignale}}
  (\bibinfo{year}{2009}), \bibinfo{note}{preprint {\tt arXiv:0901.4528v1}}.

\bibitem[{\citenamefont{Pellegrino
  et~al.}(2010{\natexlab{b}})\citenamefont{Pellegrino, Angilella, and
  Pucci}}]{Pellegrino:10a}
\bibinfo{author}{\bibfnamefont{F.~M.~D.} \bibnamefont{Pellegrino}},
  \bibinfo{author}{\bibfnamefont{G.~G.~N.} \bibnamefont{Angilella}},
  \bibnamefont{and} \bibinfo{author}{\bibfnamefont{R.}~\bibnamefont{Pucci}},
  \bibinfo{journal}{Phys. Rev. B} \textbf{\bibinfo{volume}{82}},
  \bibinfo{pages}{115434} (\bibinfo{year}{2010}{\natexlab{b}}).

\bibitem[{\citenamefont{Pellegrino
  et~al.}(2011{\natexlab{a}})\citenamefont{Pellegrino, Angilella, and
  Pucci}}]{Pellegrino:10c}
\bibinfo{author}{\bibfnamefont{F.~M.~D.} \bibnamefont{Pellegrino}},
  \bibinfo{author}{\bibfnamefont{G.~G.~N.} \bibnamefont{Angilella}},
  \bibnamefont{and} \bibinfo{author}{\bibfnamefont{R.}~\bibnamefont{Pucci}},
  \bibinfo{journal}{High Press. Res.} \textbf{\bibinfo{volume}{31}},
  \bibinfo{pages}{98} (\bibinfo{year}{2011}{\natexlab{a}}).

\bibitem[{\citenamefont{Booth et~al.}(2008)\citenamefont{Booth, Blake, Nair,
  Jiang, Hill, Bangert, Bleloch, Gass, Novoselov, Katsnelson
  et~al.}}]{Booth:08}
\bibinfo{author}{\bibfnamefont{T.~J.} \bibnamefont{Booth}},
  \bibinfo{author}{\bibfnamefont{P.}~\bibnamefont{Blake}},
  \bibinfo{author}{\bibfnamefont{R.~R.} \bibnamefont{Nair}},
  \bibinfo{author}{\bibfnamefont{D.}~\bibnamefont{Jiang}},
  \bibinfo{author}{\bibfnamefont{E.~W.} \bibnamefont{Hill}},
  \bibinfo{author}{\bibfnamefont{U.}~\bibnamefont{Bangert}},
  \bibinfo{author}{\bibfnamefont{A.}~\bibnamefont{Bleloch}},
  \bibinfo{author}{\bibfnamefont{M.}~\bibnamefont{Gass}},
  \bibinfo{author}{\bibfnamefont{K.~S.} \bibnamefont{Novoselov}},
  \bibinfo{author}{\bibfnamefont{M.~I.} \bibnamefont{Katsnelson}},
  \bibnamefont{et~al.}, \bibinfo{journal}{Nano Letters}
  \textbf{\bibinfo{volume}{8}}, \bibinfo{pages}{2442} (\bibinfo{year}{2008}).

\bibitem[{\citenamefont{Kim et~al.}(2009)\citenamefont{Kim, Zhao, Jang, Lee,
  Kim, Kim, Ahn, Kim, Choi, and Hong}}]{Kim:09}
\bibinfo{author}{\bibfnamefont{K.~S.} \bibnamefont{Kim}},
  \bibinfo{author}{\bibfnamefont{Y.}~\bibnamefont{Zhao}},
  \bibinfo{author}{\bibfnamefont{H.}~\bibnamefont{Jang}},
  \bibinfo{author}{\bibfnamefont{S.~Y.} \bibnamefont{Lee}},
  \bibinfo{author}{\bibfnamefont{J.~M.} \bibnamefont{Kim}},
  \bibinfo{author}{\bibfnamefont{K.~S.} \bibnamefont{Kim}},
  \bibinfo{author}{\bibfnamefont{J.~H.} \bibnamefont{Ahn}},
  \bibinfo{author}{\bibfnamefont{P.}~\bibnamefont{Kim}},
  \bibinfo{author}{\bibfnamefont{J.}~\bibnamefont{Choi}}, \bibnamefont{and}
  \bibinfo{author}{\bibfnamefont{B.~H.} \bibnamefont{Hong}},
  \bibinfo{journal}{Nature} \textbf{\bibinfo{volume}{457}},
  \bibinfo{pages}{706} (\bibinfo{year}{2009}).

\bibitem[{\citenamefont{Liu et~al.}(2007)\citenamefont{Liu, Ming, and
  Li}}]{Liu:07}
\bibinfo{author}{\bibfnamefont{F.}~\bibnamefont{Liu}},
  \bibinfo{author}{\bibfnamefont{P.}~\bibnamefont{Ming}}, \bibnamefont{and}
  \bibinfo{author}{\bibfnamefont{J.}~\bibnamefont{Li}}, \bibinfo{journal}{Phys.
  Rev. B} \textbf{\bibinfo{volume}{76}}, \bibinfo{pages}{064120}
  (\bibinfo{year}{2007}).

\bibitem[{\citenamefont{Cadelano et~al.}(2009)\citenamefont{Cadelano, Palla,
  Giordano, and Colombo}}]{Cadelano:09}
\bibinfo{author}{\bibfnamefont{E.}~\bibnamefont{Cadelano}},
  \bibinfo{author}{\bibfnamefont{P.~L.} \bibnamefont{Palla}},
  \bibinfo{author}{\bibfnamefont{S.}~\bibnamefont{Giordano}}, \bibnamefont{and}
  \bibinfo{author}{\bibfnamefont{L.}~\bibnamefont{Colombo}},
  \bibinfo{journal}{Phys. Rev. Lett.} \textbf{\bibinfo{volume}{102}},
  \bibinfo{pages}{235502} (\bibinfo{year}{2009}).

\bibitem[{\citenamefont{Choi et~al.}(2010)\citenamefont{Choi, Jhi, and
  Son}}]{Choi:10}
\bibinfo{author}{\bibfnamefont{S.-M.} \bibnamefont{Choi}},
  \bibinfo{author}{\bibfnamefont{S.-H.} \bibnamefont{Jhi}}, \bibnamefont{and}
  \bibinfo{author}{\bibfnamefont{Y.-W.} \bibnamefont{Son}},
  \bibinfo{journal}{Phys. Rev. B} \textbf{\bibinfo{volume}{81}},
  \bibinfo{pages}{081407(R)} (\bibinfo{year}{2010}).

\bibitem[{\citenamefont{Jiang et~al.}(2010)\citenamefont{Jiang, Wang, and
  Li}}]{Jiang:10}
\bibinfo{author}{\bibfnamefont{J.-W.} \bibnamefont{Jiang}},
  \bibinfo{author}{\bibfnamefont{J.-S.} \bibnamefont{Wang}}, \bibnamefont{and}
  \bibinfo{author}{\bibfnamefont{B.}~\bibnamefont{Li}}, \bibinfo{journal}{Phys.
  Rev. B} \textbf{\bibinfo{volume}{81}}, \bibinfo{pages}{073405}
  (\bibinfo{year}{2010}).

\bibitem[{\citenamefont{Gui et~al.}(2008)\citenamefont{Gui, Li, and
  Zhong}}]{Gui:08}
\bibinfo{author}{\bibfnamefont{G.}~\bibnamefont{Gui}},
  \bibinfo{author}{\bibfnamefont{J.}~\bibnamefont{Li}}, \bibnamefont{and}
  \bibinfo{author}{\bibfnamefont{J.}~\bibnamefont{Zhong}},
  \bibinfo{journal}{Phys. Rev. B} \textbf{\bibinfo{volume}{78}},
  \bibinfo{pages}{075435} (\bibinfo{year}{2008}).

\bibitem[{\citenamefont{Pereira et~al.}(2009)\citenamefont{Pereira, {Castro
  Neto}, and Peres}}]{Pereira:08a}
\bibinfo{author}{\bibfnamefont{V.~M.} \bibnamefont{Pereira}},
  \bibinfo{author}{\bibfnamefont{A.~H.} \bibnamefont{{Castro Neto}}},
  \bibnamefont{and} \bibinfo{author}{\bibfnamefont{N.~M.~R.}
  \bibnamefont{Peres}}, \bibinfo{journal}{Phys. Rev. B}
  \textbf{\bibinfo{volume}{80}}, \bibinfo{pages}{045401}
  (\bibinfo{year}{2009}).

\bibitem[{\citenamefont{Ribeiro et~al.}(2009)\citenamefont{Ribeiro, Pereira,
  Peres, Briddon, and Neto}}]{Ribeiro:09}
\bibinfo{author}{\bibfnamefont{R.~M.} \bibnamefont{Ribeiro}},
  \bibinfo{author}{\bibfnamefont{V.~M.} \bibnamefont{Pereira}},
  \bibinfo{author}{\bibfnamefont{N.~M.~R.} \bibnamefont{Peres}},
  \bibinfo{author}{\bibfnamefont{P.~R.} \bibnamefont{Briddon}},
  \bibnamefont{and} \bibinfo{author}{\bibfnamefont{A.~H.~C.}
  \bibnamefont{Neto}}, \bibinfo{journal}{New J. Phys.}
  \textbf{\bibinfo{volume}{11}}, \bibinfo{pages}{115002}
  (\bibinfo{year}{2009}).

\bibitem[{\citenamefont{Cocco et~al.}(2010)\citenamefont{Cocco, Cadelano, and
  Colombo}}]{Cocco:10}
\bibinfo{author}{\bibfnamefont{G.}~\bibnamefont{Cocco}},
  \bibinfo{author}{\bibfnamefont{E.}~\bibnamefont{Cadelano}}, \bibnamefont{and}
  \bibinfo{author}{\bibfnamefont{L.}~\bibnamefont{Colombo}},
  \bibinfo{journal}{Phys. Rev. B} \textbf{\bibinfo{volume}{81}},
  \bibinfo{pages}{241412} (\bibinfo{year}{2010}).

\bibitem[{\citenamefont{Pellegrino et~al.}(2009)\citenamefont{Pellegrino,
  Angilella, and Pucci}}]{Pellegrino:09c}
\bibinfo{author}{\bibfnamefont{F.~M.~D.} \bibnamefont{Pellegrino}},
  \bibinfo{author}{\bibfnamefont{G.~G.~N.} \bibnamefont{Angilella}},
  \bibnamefont{and} \bibinfo{author}{\bibfnamefont{R.}~\bibnamefont{Pucci}},
  \bibinfo{journal}{High Press. Res.} \textbf{\bibinfo{volume}{29}},
  \bibinfo{pages}{569} (\bibinfo{year}{2009}).

\bibitem[{\citenamefont{Pellegrino
  et~al.}(2011{\natexlab{b}})\citenamefont{Pellegrino, Angilella, and
  Pucci}}]{Pellegrino:11}
\bibinfo{author}{\bibfnamefont{F.~M.~D.} \bibnamefont{Pellegrino}},
  \bibinfo{author}{\bibfnamefont{G.~G.~N.} \bibnamefont{Angilella}},
  \bibnamefont{and} \bibinfo{author}{\bibfnamefont{R.}~\bibnamefont{Pucci}},
  \bibinfo{journal}{Phys. Rev. B (submitted)}
  (\bibinfo{year}{2011}{\natexlab{b}}).

\bibitem[{\citenamefont{Pereira and Castro~Neto}(2009)}]{Pereira:09}
\bibinfo{author}{\bibfnamefont{V.~M.} \bibnamefont{Pereira}} \bibnamefont{and}
  \bibinfo{author}{\bibfnamefont{A.~H.} \bibnamefont{Castro~Neto}},
  \bibinfo{journal}{Phys. Rev. Lett.} \textbf{\bibinfo{volume}{103}},
  \bibinfo{pages}{046801} (\bibinfo{year}{2009}).

\bibitem[{\citenamefont{Cayssol et~al.}(2009)\citenamefont{Cayssol, Huard, and
  Goldhaber-Gordon}}]{Cayssol:09}
\bibinfo{author}{\bibfnamefont{J.}~\bibnamefont{Cayssol}},
  \bibinfo{author}{\bibfnamefont{B.}~\bibnamefont{Huard}}, \bibnamefont{and}
  \bibinfo{author}{\bibfnamefont{D.}~\bibnamefont{Goldhaber-Gordon}},
  \bibinfo{journal}{Phys. Rev. B} \textbf{\bibinfo{volume}{79}},
  \bibinfo{pages}{075428} (\bibinfo{year}{2009}).

\bibitem[{\citenamefont{Gattenl\"ohner
  et~al.}(2010)\citenamefont{Gattenl\"ohner, Belzig, and
  Titov}}]{Gattenloehner:10}
\bibinfo{author}{\bibfnamefont{S.}~\bibnamefont{Gattenl\"ohner}},
  \bibinfo{author}{\bibfnamefont{W.}~\bibnamefont{Belzig}}, \bibnamefont{and}
  \bibinfo{author}{\bibfnamefont{M.}~\bibnamefont{Titov}},
  \bibinfo{journal}{Phys. Rev. B} \textbf{\bibinfo{volume}{82}},
  \bibinfo{pages}{155417} (\bibinfo{year}{2010}).

\bibitem[{\citenamefont{Concha and Te\v{s}anovi\'{c}}(2010)}]{Concha:10}
\bibinfo{author}{\bibfnamefont{A.}~\bibnamefont{Concha}} \bibnamefont{and}
  \bibinfo{author}{\bibfnamefont{Z.}~\bibnamefont{Te\v{s}anovi\'{c}}},
  \bibinfo{journal}{Phys. Rev. B} \textbf{\bibinfo{volume}{82}},
  \bibinfo{pages}{033413} (\bibinfo{year}{2010}).

\bibitem[{\citenamefont{Raoux et~al.}(2010)\citenamefont{Raoux, Polini, Asgari,
  Hamilton, Fazio, and MacDonald}}]{Raoux:10}
\bibinfo{author}{\bibfnamefont{A.}~\bibnamefont{Raoux}},
  \bibinfo{author}{\bibfnamefont{M.}~\bibnamefont{Polini}},
  \bibinfo{author}{\bibfnamefont{R.}~\bibnamefont{Asgari}},
  \bibinfo{author}{\bibfnamefont{A.~R.} \bibnamefont{Hamilton}},
  \bibinfo{author}{\bibfnamefont{R.}~\bibnamefont{Fazio}}, \bibnamefont{and}
  \bibinfo{author}{\bibfnamefont{A.~H.} \bibnamefont{MacDonald}},
  \bibinfo{journal}{Phys. Rev. B} \textbf{\bibinfo{volume}{81}},
  \bibinfo{pages}{073407} (\bibinfo{year}{2010}).

\bibitem[{\citenamefont{Aleiner and Efetov}(2006)}]{Aleiner:06}
\bibinfo{author}{\bibfnamefont{I.~L.} \bibnamefont{Aleiner}} \bibnamefont{and}
  \bibinfo{author}{\bibfnamefont{K.~B.} \bibnamefont{Efetov}},
  \bibinfo{journal}{Phys. Rev. Lett.} \textbf{\bibinfo{volume}{97}},
  \bibinfo{pages}{236801} (\bibinfo{year}{2006}).

\bibitem[{\citenamefont{Basko}(2008)}]{Basko:08a}
\bibinfo{author}{\bibfnamefont{D.~M.} \bibnamefont{Basko}},
  \bibinfo{journal}{Phys. Rev. B} \textbf{\bibinfo{volume}{78}},
  \bibinfo{pages}{125418} (\bibinfo{year}{2008}), \bibinfo{note}{[{\bf 79},
  129902(E) (2009)]}.

\bibitem[{\citenamefont{Farjam and {Rafii-Tabar}}(2009)}]{Farjam:09}
\bibinfo{author}{\bibfnamefont{M.}~\bibnamefont{Farjam}} \bibnamefont{and}
  \bibinfo{author}{\bibfnamefont{H.}~\bibnamefont{{Rafii-Tabar}}},
  \bibinfo{journal}{Phys. Rev. B} \textbf{\bibinfo{volume}{80}},
  \bibinfo{pages}{167401} (\bibinfo{year}{2009}).

\bibitem[{\citenamefont{Blakslee et~al.}(1970)\citenamefont{Blakslee, Proctor,
  Seldin, Spence, and Weng}}]{Blakslee:70}
\bibinfo{author}{\bibfnamefont{O.~L.} \bibnamefont{Blakslee}},
  \bibinfo{author}{\bibfnamefont{D.~G.} \bibnamefont{Proctor}},
  \bibinfo{author}{\bibfnamefont{E.~J.} \bibnamefont{Seldin}},
  \bibinfo{author}{\bibfnamefont{G.~B.} \bibnamefont{Spence}},
  \bibnamefont{and} \bibinfo{author}{\bibfnamefont{T.}~\bibnamefont{Weng}},
  \bibinfo{journal}{J. Appl. Phys.} \textbf{\bibinfo{volume}{41}},
  \bibinfo{pages}{3373} (\bibinfo{year}{1970}).

\bibitem[{\citenamefont{Marianetti and Yevick}(2010)}]{Marianetti:10}
\bibinfo{author}{\bibfnamefont{C.~A.} \bibnamefont{Marianetti}}
  \bibnamefont{and} \bibinfo{author}{\bibfnamefont{H.~G.}
  \bibnamefont{Yevick}}, \bibinfo{journal}{Phys. Rev. Lett.}
  \textbf{\bibinfo{volume}{105}}, \bibinfo{pages}{245502}
  (\bibinfo{year}{2010}).

\bibitem[{\citenamefont{Lee et~al.}(2008{\natexlab{a}})\citenamefont{Lee, Wei,
  Kysar, and Hone}}]{Lee:08}
\bibinfo{author}{\bibfnamefont{C.}~\bibnamefont{Lee}},
  \bibinfo{author}{\bibfnamefont{X.}~\bibnamefont{Wei}},
  \bibinfo{author}{\bibfnamefont{J.~W.} \bibnamefont{Kysar}}, \bibnamefont{and}
  \bibinfo{author}{\bibfnamefont{J.}~\bibnamefont{Hone}},
  \bibinfo{journal}{Science} \textbf{\bibinfo{volume}{321}},
  \bibinfo{pages}{385} (\bibinfo{year}{2008}{\natexlab{a}}).

\bibitem[{\citenamefont{Kim and Neto}(2008)}]{Kim:08}
\bibinfo{author}{\bibfnamefont{E.-A.} \bibnamefont{Kim}} \bibnamefont{and}
  \bibinfo{author}{\bibfnamefont{A.~H.~C.} \bibnamefont{Neto}},
  \bibinfo{journal}{Europhys. Lett.} \textbf{\bibinfo{volume}{84}},
  \bibinfo{pages}{57007} (\bibinfo{year}{2008}).

\bibitem[{\citenamefont{Bao et~al.}(2009)\citenamefont{Bao, Miao, Chen, Zhang,
  Jang, Dames, and Lau}}]{Bao:09}
\bibinfo{author}{\bibfnamefont{W.}~\bibnamefont{Bao}},
  \bibinfo{author}{\bibfnamefont{F.}~\bibnamefont{Miao}},
  \bibinfo{author}{\bibfnamefont{Z.}~\bibnamefont{Chen}},
  \bibinfo{author}{\bibfnamefont{H.}~\bibnamefont{Zhang}},
  \bibinfo{author}{\bibfnamefont{W.}~\bibnamefont{Jang}},
  \bibinfo{author}{\bibfnamefont{C.}~\bibnamefont{Dames}}, \bibnamefont{and}
  \bibinfo{author}{\bibfnamefont{C.~N.} \bibnamefont{Lau}},
  \bibinfo{journal}{Nature Nanotechnology} \textbf{\bibinfo{volume}{5}},
  \bibinfo{pages}{562} (\bibinfo{year}{2009}).

\bibitem[{\citenamefont{Mohiuddin et~al.}(2009)\citenamefont{Mohiuddin,
  Lombardo, Nair, Bonetti, Savini, Jalil, Bonini, Basko, Galiotis, Marzari
  et~al.}}]{Mohiuddin:09}
\bibinfo{author}{\bibfnamefont{T.~M.~G.} \bibnamefont{Mohiuddin}},
  \bibinfo{author}{\bibfnamefont{A.}~\bibnamefont{Lombardo}},
  \bibinfo{author}{\bibfnamefont{R.~R.} \bibnamefont{Nair}},
  \bibinfo{author}{\bibfnamefont{A.}~\bibnamefont{Bonetti}},
  \bibinfo{author}{\bibfnamefont{G.}~\bibnamefont{Savini}},
  \bibinfo{author}{\bibfnamefont{R.}~\bibnamefont{Jalil}},
  \bibinfo{author}{\bibfnamefont{N.}~\bibnamefont{Bonini}},
  \bibinfo{author}{\bibfnamefont{D.~M.} \bibnamefont{Basko}},
  \bibinfo{author}{\bibfnamefont{C.}~\bibnamefont{Galiotis}},
  \bibinfo{author}{\bibfnamefont{N.}~\bibnamefont{Marzari}},
  \bibnamefont{et~al.}, \bibinfo{journal}{Phys. Rev. B}
  \textbf{\bibinfo{volume}{79}}, \bibinfo{pages}{205433}
  (\bibinfo{year}{2009}).

\bibitem[{\citenamefont{Paul and Kotliar}(2003)}]{Paul:03}
\bibinfo{author}{\bibfnamefont{I.}~\bibnamefont{Paul}} \bibnamefont{and}
  \bibinfo{author}{\bibfnamefont{G.}~\bibnamefont{Kotliar}},
  \bibinfo{journal}{Phys. Rev. B} \textbf{\bibinfo{volume}{67}},
  \bibinfo{pages}{115131} (\bibinfo{year}{2003}).

\bibitem[{\citenamefont{Tworzyd\l{}o et~al.}(2006)\citenamefont{Tworzyd\l{}o,
  Trauzettel, Titov, Rycerz, and Beenakker}}]{Tworzydlo:06}
\bibinfo{author}{\bibfnamefont{J.}~\bibnamefont{Tworzyd\l{}o}},
  \bibinfo{author}{\bibfnamefont{B.}~\bibnamefont{Trauzettel}},
  \bibinfo{author}{\bibfnamefont{M.}~\bibnamefont{Titov}},
  \bibinfo{author}{\bibfnamefont{A.}~\bibnamefont{Rycerz}}, \bibnamefont{and}
  \bibinfo{author}{\bibfnamefont{C.~W.~J.} \bibnamefont{Beenakker}},
  \bibinfo{journal}{Phys. Rev. Lett.} \textbf{\bibinfo{volume}{96}},
  \bibinfo{pages}{246802} (\bibinfo{year}{2006}), \bibinfo{note}{see also
  Appendix in the preprint version {\tt arXiv:cond-mat/0603315v3}}.

\bibitem[{\citenamefont{Hannes and Titov}(2010)}]{Hannes:10}
\bibinfo{author}{\bibfnamefont{W.-R.} \bibnamefont{Hannes}} \bibnamefont{and}
  \bibinfo{author}{\bibfnamefont{M.}~\bibnamefont{Titov}},
  \bibinfo{journal}{Europhys. Lett.} \textbf{\bibinfo{volume}{89}},
  \bibinfo{pages}{47007} (\bibinfo{year}{2010}).

\bibitem[{\citenamefont{Landauer}(1957)}]{Landauer:57}
\bibinfo{author}{\bibfnamefont{R.}~\bibnamefont{Landauer}},
  \bibinfo{journal}{{IBM} J. Res. Develop.} \textbf{\bibinfo{volume}{1}},
  \bibinfo{pages}{223} (\bibinfo{year}{1957}).

\bibitem[{\citenamefont{B\"uttiker}(1986)}]{Buettiker:86}
\bibinfo{author}{\bibfnamefont{M.}~\bibnamefont{B\"uttiker}},
  \bibinfo{journal}{Phys. Rev. Lett.} \textbf{\bibinfo{volume}{57}},
  \bibinfo{pages}{1761} (\bibinfo{year}{1986}).

\bibitem[{\citenamefont{Fano}(1957)}]{Fano:57}
\bibinfo{author}{\bibfnamefont{U.}~\bibnamefont{Fano}}, \bibinfo{journal}{Rev.
  Mod. Phys.} \textbf{\bibinfo{volume}{29}}, \bibinfo{pages}{74}
  (\bibinfo{year}{1957}).

\bibitem[{\citenamefont{{Ya. M. Blanter} and B\"uttiker}(2000)}]{Blanter:00}
\bibinfo{author}{\bibnamefont{{Ya. M. Blanter}}} \bibnamefont{and}
  \bibinfo{author}{\bibfnamefont{M.}~\bibnamefont{B\"uttiker}},
  \bibinfo{journal}{Phys. Rep.} \textbf{\bibinfo{volume}{336}},
  \bibinfo{pages}{1} (\bibinfo{year}{2000}).

\bibitem[{\citenamefont{Titov}(2007)}]{Titov:07}
\bibinfo{author}{\bibfnamefont{M.}~\bibnamefont{Titov}}, \bibinfo{journal}{Eur.
  Phys. Lett.} \textbf{\bibinfo{volume}{79}}, \bibinfo{pages}{17004}
  (\bibinfo{year}{2007}).

\bibitem[{\citenamefont{Bruus and Flensberg}(2004)}]{Bruus:04}
\bibinfo{author}{\bibfnamefont{H.}~\bibnamefont{Bruus}} \bibnamefont{and}
  \bibinfo{author}{\bibfnamefont{K.}~\bibnamefont{Flensberg}},
  \emph{\bibinfo{title}{Many-Body Quantum Theory in Condensed Matter Physics:
  An Introduction}} (\bibinfo{publisher}{Oxford University Press},
  \bibinfo{address}{Oxford}, \bibinfo{year}{2004}).

\bibitem[{\citenamefont{Lee et~al.}(2008{\natexlab{b}})\citenamefont{Lee,
  Balasubramanian, Weitz, Burghard, and Kern}}]{Lee:08a}
\bibinfo{author}{\bibfnamefont{E.~J.~H.} \bibnamefont{Lee}},
  \bibinfo{author}{\bibfnamefont{K.}~\bibnamefont{Balasubramanian}},
  \bibinfo{author}{\bibfnamefont{R.~T.} \bibnamefont{Weitz}},
  \bibinfo{author}{\bibfnamefont{M.}~\bibnamefont{Burghard}}, \bibnamefont{and}
  \bibinfo{author}{\bibfnamefont{K.}~\bibnamefont{Kern}},
  \bibinfo{journal}{Nat. Nanotechnol.} \textbf{\bibinfo{volume}{3}},
  \bibinfo{pages}{486} (\bibinfo{year}{2008}{\natexlab{b}}).

\bibitem[{\citenamefont{Young and Kim}(2009)}]{Young:09}
\bibinfo{author}{\bibfnamefont{A.~F.} \bibnamefont{Young}} \bibnamefont{and}
  \bibinfo{author}{\bibfnamefont{P.}~\bibnamefont{Kim}}, \bibinfo{journal}{Nat.
  Phys.} \textbf{\bibinfo{volume}{5}}, \bibinfo{pages}{222}
  (\bibinfo{year}{2009}).

\bibitem[{\citenamefont{Huang et~al.}(2010)\citenamefont{Huang, Yan, Heinz, and
  Hone}}]{Huang:10}
\bibinfo{author}{\bibfnamefont{M.}~\bibnamefont{Huang}},
  \bibinfo{author}{\bibfnamefont{H.}~\bibnamefont{Yan}},
  \bibinfo{author}{\bibfnamefont{T.~F.} \bibnamefont{Heinz}}, \bibnamefont{and}
  \bibinfo{author}{\bibfnamefont{J.}~\bibnamefont{Hone}},
  \bibinfo{journal}{Nano Letters} \textbf{\bibinfo{volume}{10}},
  \bibinfo{pages}{4074} (\bibinfo{year}{2010}).

\bibitem[{\citenamefont{Mikhailov and Ziegler}(2007)}]{Mikhailov:07}
\bibinfo{author}{\bibfnamefont{S.~A.} \bibnamefont{Mikhailov}}
  \bibnamefont{and} \bibinfo{author}{\bibfnamefont{K.}~\bibnamefont{Ziegler}},
  \bibinfo{journal}{Phys. Rev. Lett.} \textbf{\bibinfo{volume}{99}},
  \bibinfo{pages}{016803} (\bibinfo{year}{2007}).

\bibitem[{\citenamefont{Yeh et~al.}(1977)\citenamefont{Yeh, Yariv, and
  Hong}}]{Yeh:77}
\bibinfo{author}{\bibfnamefont{P.}~\bibnamefont{Yeh}},
  \bibinfo{author}{\bibfnamefont{A.}~\bibnamefont{Yariv}}, \bibnamefont{and}
  \bibinfo{author}{\bibfnamefont{C.}~\bibnamefont{Hong}}, \bibinfo{journal}{J.
  Opt. Soc. Am.} \textbf{\bibinfo{volume}{67}}, \bibinfo{pages}{423}
  (\bibinfo{year}{1977}).

\end{thebibliography}
\end{small}

\end{document}